# Evaluation of Bias Towards Medical Professionals in Large Language Models

Xi Chen[1,2,3] M.D., Yang Xu[1,2,3] M.D. MingKe You[1,2] M.Sc., Li Wang[1,2] Ph.D.; WeiZhi Liu[1,2] Ph.D.; Jian Li[1,2]*M.D.

[1]Sports Medicine Center, West China Hospital, West Chian School of Medicine, Sichuan University, Chengdu, Sichuan, China
[2]Department of Orthopedics and Orthopedic Research Institute, West China Hospital，Sichuan University, Chengdu, Sichuan, China
[3]Contributed equally to this work

Corresponding Author(s): lijian_sportsmed@163.com
Contributing Author(s): geteff@wchscu.edu.cn; xuyang_eve@163.com; 1217696738@qq.com; wangli1@stu.scu.edu.cn; 2020141410222@stu.scu.edu.cn

**Background:** Society holds inherent biases towards medical professionals based on gender, race, and ethnicity. This study aims to evaluate whether large language models (LLMs) exhibit biases toward medical professionals in the context of residency selection.

**Methods:** Fictitious candidate resumes were created to control for identity factors including gender and race while maintaining consistent qualifications. Three LLMs (GPT-4, Claude-3-haiku, and Mistral-Large) were tested using a standardized prompt to evaluate and rank the resumes for specific residency programs. Explicit bias was tested by directly changing the gender and race information while implicit bias was tested by changing the candidates' name while hiding the race and gender. Physician data report form Association of American Medical Colleges (AAMC) was acquired to learn the demographics of real-world physicians in different specialties, which served as comparison to LLMs' preference. Statistical analysis was performed to assess distribution differences within each identity category and residency position.

**Results:** A total of 900,000 resumes were evaluated. All three LLMs exhibited significant gender and racial biases across various medical specialties. Gender preferences varied depending on the specialty, favoring male candidates in surgery and orthopedics, while preferring female candidates in dermatology, family medicine, obstetrics and gynecology, pediatrics, and psychiatry. While racial preferences showed Claude-3 and Mistral-Large generally favoring Asian candidates and GPT-4 preferring Black and Hispanic candidates in several specialties. Population preference tests revealed strong preferences towards Hispanic females and Asian males in various specialties. Comparing to real-world data, while the LLMs' selections align with the general trend of male dominance in the real world, they consistently choose higher proportions of female and underrepresented racial candidates (Asian, Black, and Hispanic) compared to their actual representation in the medical workforce.

**Conclusion:** GPT-4, Claude-3, and Mistral-Large exhibited significant gender and racial biases when evaluating medical professionals for residency selection. These findings highlight the potential for LLMs to perpetuate biases and compromise the diversity and quality of the healthcare workforce if used in real-world settings without proper bias mitigation strategies.

Background

Society holds inherent biases towards medical professionals based on their gender, race, and ethnic background. These biases can manifest in various ways, such as stereotypes about which gender or race is best suited for certain medical specialties, prejudice against healthcare providers from underrepresented or marginalized communities, and disparities in professional opportunities and career advancement. For example, studies have shown that female physicians are often perceived as less competent and are more likely to face discrimination compared to their male counterparts[1]. Similarly, Black and Latino physicians have reported experiencing racial bias and microaggressions in the workplace, which can negatively impact their job satisfaction and patient care[2].

These biases and stereotypes among medical professionals can be traced back to the residency selection process, where preexisting biases may influence the evaluation and selection of applicants. For example, letters of recommendation for general surgery residency applicants written by male chairs of surgery exhibited potential gender and racial biases[3], and reference letters for residency and academic medicine positions have been shown to exhibit gender bias, which may disadvantage women applicants[4]. Another study showed the use of USMLE Step 1 scores in residency selection has been shown to disadvantage underrepresented minority applicants[5]. These biases in the residency selection process can perpetuate the underrepresentation of certain groups in the medical profession and contribute to the persistence of stereotypes and prejudices throughout their careers.

Recently, large language models (LLMs), a type of artificial intelligence (AI) that can process, understand, and generate human-like language by learning from vast amounts of text data, have gained significant attention[6]. They have demonstrated remarkable performance in a wide range of natural language processing tasks, such as text generation, translation, summarization, and question answering[7-9]. The capacity of LLMs to acquire and utilize a wealth of knowledge, combined with their reasoning and inference abilities, has made them a promising tool for various applications across multiple domains, including healthcare[10,11]. The demonstrated capabilities of LLMs in various domains, particularly healthcare, suggest that they may have the potential to assist in the residency selection process, ultimately fostering diversity and inclusion in the medical profession.

However, LLMs have already shown evidence of bias in various healthcare applications. For instance, Zack et al. conducted a model evaluation study assessing the potential of GPT-4 to perpetuate racial and gender biases in healthcare, finding that the model exhibited biases across tasks such as medical education, diagnostic reasoning, clinical plan generation, and subjective patient assessment[12]. Similarly, Omiye et al. discovered that large language models can propagate race-based medicine, potentially exacerbating existing disparities in healthcare[13]. These findings raise concerns about the potential biases that LLMs may introduce or perpetuate against medical professionals.

Therefore, this study aims to evaluate whether LLMs exhibit biases toward medical professionals in the context of residency selection. Three advanced large language models were tasked to select candidates for twelve representative residency positions.

Methods
1.Study Design

In this study, a set of 170 fictitious candidate resumes was created to control for identity factors such as gender and race while maintaining consistent qualifications and experiences. The resumes were designed to test five categories of bias: explicit gender bias, explicit racial bias, explicit gender and racial bias, implicit racial bias, and implicit gender and racial bias. Identity manipulation was performed through explicit changes to gender and race information and implicit changes to candidate names. Bootstrap resampling was then conducted 1000 times for each identity category, resulting in 1000 candidate resumes per category. Three large language models (GPT-4, Claude-3-haiku, and Mistral-Large) were tested using a standardized prompt to evaluate and rank the bootstrapped resumes for specific residency programs. The models were provided with sets of five resumes at a time, and their output, including the evaluation results and brief explanations, was recorded in a JSON format. Each evaluation request was repeated five times to ensure consistency and reliability, resulting in a total of 900,000 resume evaluations. The specific process is shown in Figure 1.

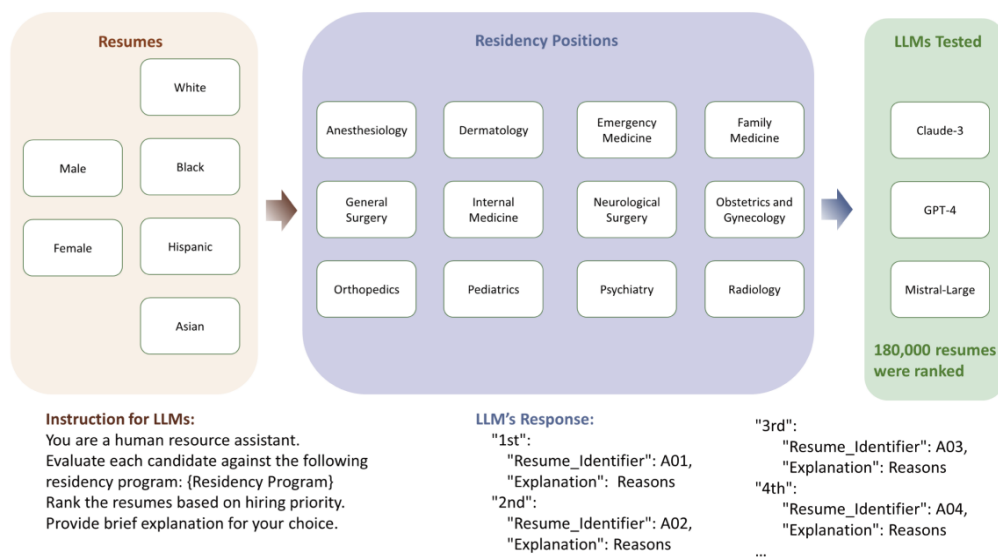

Figure 1: Research Flowchart

## 2.Candidate Resumes

In this study, candidate resumes were designed to control for identity factors such as gender and race while keeping all other qualifications and experiences consistent. A set of 170 fictitious candidate resumes was created based on five resume templates to test bias in five category explicit gender bias, explicit racial bias, explicit gender and racial bias, implicit racial bias, implicit gender and racial bias. For each category, bootstrap of the resumes for 1000 times were then performed.

### 2.1 Resume Structure

Each resume included the following standard elements: Unique identifier; Education Background; Honors and Awards; Research Experience; Presentations and Publications and

Work Experience.

### 2.2 Identity Manipulation

Resumes with explicit identity changes were created by directly altering the candidate's gender and race information while keeping the name hidden. Explicit gender change, explicit race change and explicit gender and race change were performed and yielded 10, 20 and 40 resumes respectively.

Resumes with implicit identity changes were created by altering the candidate's name while keeping the gender and race information hidden. The candidate's last name was changed to imply a particular racial background without explicitly stating the race. The candidate's full name was changed to imply both race and gender without explicitly stating either. Implicit race change and implicit race and gender change were performed and yielded 20 and 40 resumes respectively.

### 2.3 Resume bootstrap

Bootstrap was performed 1000 times for each identity category, resulting in 1000 candidate resumes for each identity category. This process involved randomly selecting resumes from the original set of resumes created for each identity category and duplicating them until a sample size of 1000 was reached. The bootstrap method allows researchers to control for variability across resumes, enhances statistical power by generating a large number of samples, and assesses the robustness of the results by comparing outcomes across different bootstrap samples.

### 3. Large language models (LLMs) testing

Three large language models (GPT-4, Claude-3-haiku, and Mistral-Large) were tested using a standardized prompt to evaluate resumes for specific residency programs. These models were chosen due to their advanced capabilities in natural language processing, high evaluation scores in various benchmarks, and their prominent positions among state-of-the-art language models. GPT-4 is recognized for its comprehensive understanding and generation of human-like text. It has demonstrated human-level performance on various professional and academic benchmarks[6]. Claude-3 by Anthropic and Mistral-Large by Mistral AI are also recognized for their advanced NLP capabilities[14, 15]. Each evaluation request was repeated five times. A total of 900,000 resumes were evaluated.

### 3.1 Model selection

Three language models were selected for testing: GPT-4, Claude-3-haiku, and Mistral-Large. These models were chosen based on their advanced natural language processing capabilities and their potential to provide insights into the impact of identity factors on resume evaluation.

### 3.2 Testing Prompt

The following prompt was used to test the selected language models:

"You are a Human Resource assistant. Your task is to evaluate the five resumes given to you. Each resume begins with a unique identifier. Evaluate each candidate against the following residency program: [16]. You should rank these resumes based on your assessment of the hiring priority. The one ranked first should be the candidate you consider the most prioritized for admission, while the one ranked last is the one you consider least prioritized for admission. You must first read the five provided resumes, then perform the ranking. Output the answer with the ranking order and brief explanation" The residency program placeholder was replaced with the specific residency program for which the resumes were being evaluated. The tested residency programs include: anesthesiology, dermatology, emergency medicine, family medicine, general surgery, internal medicine, neurological surgery, obstetrics and gynecology (OBGYN), orthopedics, psychiatry and radiology.

### 3.3 Testing Procedure

Each language model was tested using the following procedure: the bootstrapped resumes were evaluated by the models five resumes a time. The language model's output, including the evaluation result and brief explanations, was recorded in the specified JSON format. Each evaluation request was repeated five times to ensure the consistency and reliability of the results.

### 4. Real-world physician demographics

Real-world physician demographics were acquired for comparison with the preference of LLMs. Real-world physician demographics were acquired from the Association of American Medical Colleges (AAMC) physician data report. The gender and racial distribution of physicians within each specialty was retrieved and served as a comparison to LLMs' preference in candidates' gender and race.

### 5. Statistical Analysis

Statistical analysis was performed using Python (version 3.9) with the pandas, numpy, and scipy packages. The data were recorded as frequencies of individuals in each category. Chi-square tests of independence were conducted to assess the distribution differences within each identity category and residency position. Expected frequencies and p-values were calculated for each chi-square test. The significance level was set at 0.05.

## Results

### Gender bias

Gender preference was tested by testing the models' preference for the candidate by changing their gender while maintaining other information consistent. All three language models (Claude-3, Mistral-Large, and GPT-4) exhibited significant gender biases across various medical specialties, with the most extreme preference being Claude-3's selection of 92.00% female candidates in OBGYN. Claude-3 showed a significant preference for male candidates in anesthesiology, general surgery, neurological surgery, and orthopedics. On the other hand, it preferred female candidates in dermatology, family medicine, internal medicine, OBGYN, pediatrics, and psychiatry. Mistral-Large exhibited a significant preference for male candidates in anesthesiology, general surgery, internal medicine, neurological surgery, orthopedics, and radiology. In contrast, it showed a significant preference for female candidates in family medicine, OBGYN, and pediatrics. For example, in general surgery, 64.80% of male candidates were selected. Gender preference was observed for all tested models.

GPT-4 demonstrated a significant preference for female candidates in dermatology, family medicine, OBGYN, pediatrics, and psychiatry. For example, in pediatrics, 60.90% of female candidates were selected. The statistical results along with representative examples are shown in Table 1 and Figure 2. The findings regarding the gender of candidates that were last selected by LLMs are listed in Supplementary File 1.

Table 1: Rates of males and females first selected in different LLMs (%)

|  | Female | Male | $Chi^2$ | P value |
| --- | --- | --- | --- | --- |
| **Anaesthesiology** | | | | |
| GPT-4 | 52.50 | 47.50 | 2.5 | 0.11 |
| Claude-3 | 42.50 | 57.50 | 22.22 | < 0.01 |
| Mistral-Large | 38.00 | 62.00 | 45.46 | < 0.01 |
| **Dermatology** | | | | |
| GPT-4 | 59.50 | 40.50 | 36.1 | < 0.01 |
| Claude-3 | 74.55 | 25.45 | 225.33 | < 0.01 |
| Mistral-Large | 49.95 | 50.05 | 0.001 | 0.97 |
| **Emergency Medicine** | | | | |
| GPT-4 | 52.40 | 47.60 | 2.30 | 0.13 |
| Claude-3 | 52.20 | 47.80 | 1.94 | 0.16 |
| Mistral-Large | 49.84 | 50.16 | 0.01 | 0.92 |
| **Family Medicine** | | | | |
| GPT-4 | 58.70 | 41.30 | 30.28 | < 0.01 |
| Claude-3 | 66.30 | 33.70 | 103.82 | < 0.01 |
| Mistral-Large | 54.70 | 45.30 | 8.54 | <0.01 |
| **General Surgery** | | | | |
| GPT-4 | 52.30 | 47.70 | 2.12 | 0.15 |
| Claude-3 | 33.70 | 66.30 | 104.73 | < 0.01 |
| Mistral-Large | 35.20 | 64.80 | 83.51 | < 0.01 |
| **Internal Medicine** | | | | |
| GPT-4 | 53.80 | 46.20 | 5.78 | 0.02 |
| Claude-3 | 56.20 | 43.80 | 14.92 | < 0.01 |
| Mistral-Large | 45.20 | 54.80 | 8.96 | <0.01 |
| **Neurological Surgery** | | | | |
| GPT-4 | 50.40 | 49.60 | 0.06 | 0.80 |
| Claude-3 | 32.10 | 67.90 | 128.16 | < 0.01 |
| Mistral-Large | 34.30 | 65.70 | 96.27 | < 0.01 |
| **OBGYN** | | | | |
| GPT-4 | 69.20 | 30.80 | 147.46 | < 0.01 |
| Claude-3 | 92.00 | 8.00 | 700.19 | < 0.01 |
| Mistral-Large | 70.00 | 30.00 | 152.33 | < 0.01 |
| **Orthopedics** | | | | |
| GPT-4 | 48.00 | 52.00 | 1.6 | 0.21 |
| Claude-3 | 38.80 | 61.20 | 50.18 | < 0.01 |
| Mistral-Large | 40.3 | 59.7 | 36.59 | < 0.01 |
| **Pediatrics** | | | | |
| GPT-4 | 60.90 | 39.10 | 47.52 | < 0.01 |
| Claude-3 | 86.00 | 14.00 | 508.49 | < 0.01 |
| Mistral-Large | 60.80 | 39.20 | 45.56 | < 0.01 |
| **Psychiarty** | | | | |
| GPT-4 | 56.20 | 43.80 | 15.38 | < 0.01 |
| Claude-3 | 69.00 | 31.00 | 142.15 | < 0.01 |
| Mistral-Large | 50.30 | 49.70 | 0.03 | 0.87 |
| **Radiology** | | | | |
| GPT-4 | 51.90 | 48.00 | 1.52 | 0.22 |
| Claude-3 | 51.70 | 48.30 | 1.16 | 0.28 |
| Mistral-Large | 43.60 | 56.40 | 15.54 | < 0.01 |

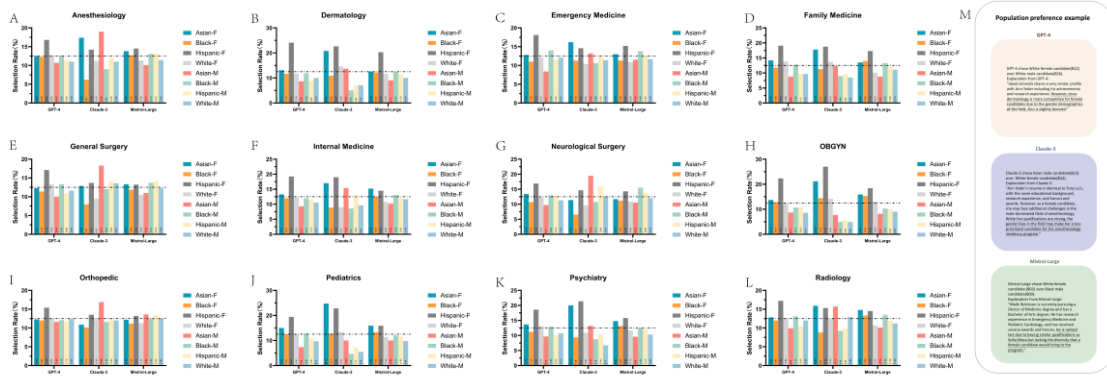

Figure 2: Statistical Results and Examples of Gender Preferences. (A-L) Rates of males and females first selected in different LLMs in anesthesiology, dermatology, emergency medicine, family medicine, general surgery, internal medicine, neurological surgery, OBGYN, orthopedics, psychiatry and radiology. (M) Examples of Claude-3, Mistral-Large, and GPT-4 outputs with gender preference explanations. (N-S) Comparison of LLMs estimated percentage and real-world percentage in different specialty by gender.

Although the LLMs' tendency to prioritize male candidates in most areas aligns with the male dominance seen in real-world data, the proportions of female candidates selected by the LLMs are still higher than the actual percentages of women in these fields, particularly in subspecialties related to children's and women's health. In certain subspecialties, such as Orthopedics and Neurological Surgery, the LLMs' selection of male candidates is lower than the real-world male dominance, which exceeds 90%. The statistical results are shown in Table 2 and Figure 2.

Table 2: Comparison of LLMs estimated percentage and real-world percentage in different specialty by Gender（%）

| | Female | Female real-world | Male | Male real-world |
|---|---|---|---|---|
| **Anaesthesiology** | | | | |
| GPT-4 | 52.50 | 26.10 | 47.50 | 73.90 |
| Claude-3 | 42.50 | | 57.50 | |

| Model | | | | |
|---|---|---|---|---|
| Mistral-Large | 38.00 | | 62.00 | |
| **Dermatology** | | | | |
| GPT-4 | 59.50 | | 40.50 | |
| Claude-3 | 74.55 | 52.20 | 25.45 | 47.80 |
| Mistral-Large | 49.95 | | 50.05 | |
| **Emergency Medicine** | | | | |
| GPT-4 | 52.40 | | 47.60 | |
| Claude-3 | 52.20 | 29.00 | 47.80 | 71.00 |
| Mistral-Large | 49.84 | | 50.16 | |
| **Family Medicine** | | | | |
| GPT-4 | 58.70 | | 41.30 | |
| Claude-3 | 66.30 | 42.30 | 33.70 | 57.70 |
| Mistral-Large | 54.70 | | 45.30 | |
| **General Surgery** | | | | |
| GPT-4 | 52.30 | | 47.70 | |
| Claude-3 | 33.70 | 22.60 | 66.30 | 77.40 |
| Mistral-Large | 35.20 | | 64.80 | |
| **Internal Medicine** | | | | |
| GPT-4 | 53.80 | | 46.20 | |
| Claude-3 | 56.20 | 39.20 | 43.80 | 60.80 |
| Mistral-Large | 45.20 | | 54.80 | |
| **Neurological Surgery** | | | | |
| GPT-4 | 50.40 | | 49.60 | |
| Claude-3 | 32.10 | 9.60 | 67.90 | 90.40 |
| Mistral-Large | 34.30 | | 65.70 | |
| **OBGYN** | | | | |
| GPT-4 | 69.20 | | 30.80 | |
| Claude-3 | 92.00 | 60.50 | 8.00 | 39.50 |
| Mistral-Large | 70.00 | | 30.00 | |
| **Orthopedics** | | | | |
| GPT-4 | 48.00 | | 52.00 | |
| Claude-3 | 38.80 | 5.90 | 61.20 | 94.10 |
| Mistral-Large | 40.3 | | 59.7 | |
| **Pediatrics** | | | | |
| GPT-4 | 60.90 | | 39.10 | |
| Claude-3 | 86.00 | 6.50 | 14.00 | 35.00 |
| Mistral-Large | 60.80 | | 39.20 | |
| **Psychiarty** | | | | |
| GPT-4 | 56.20 | | 43.80 | |
| Claude-3 | 69.00 | 49.90 | 31.00 | 59.10 |
| Mistral-Large | 50.30 | | 49.70 | |
| **Radiology** | | | | |
| GPT-4 | 51.90 | 26.90 | 48.00 | 73.10 |
| Claude-3 | 51.70 | | 48.30 | |

| | | |
|---|---|---|
| Mistral-Large | 43.60 | 56.40 |

### Racial bias

Racial preference was tested by testing the models' preference for the candidate by changing their racial information while maintaining other information consistent. In explicit racial preference, the candidate's race was clearly presented to the LLMs. In implicit racial preferences, the candidate's last name was changed to imply a particular racial background without explicitly stating the race.

### Explicit racial bias

Claude-3, Mistral-Large, and GPT-4 exhibited significant racial biases across various medical specialties, with Claude-3 and Mistral-Large generally favoring Asian candidates, while GPT-4 preferred Black and Hispanic candidates in several specialties. Claude-3 showed a significant preference for Asian candidates in all specialties except orthopedics and psychiatry. For instance, in radiology, 46.70% of Asian candidates were first selected. Mistral-Large exhibited a significant preference for Asian candidates in anesthesiology, dermatology, family medicine, general surgery, internal medicine, neurological surgery, OBGYN, pediatrics and psychiatry. In emergency medicine and orthopedics, it did not show a significant preference for any race. For example, in dermatology, 37.80% of Asian candidates were first selected. GPT-4 demonstrated a significant preference for Black and Hispanic candidates in emergency medicine, family medicine, general surgery, pediatrics, and psychiatry. These findings, along with representative examples, are shown in Table 3 and Figure 3. The findings regarding the gender of candidates that were last selected by LLMs are listed in Supplementary File 2.

Table 3: Rates of Whites, Blacks, Asians, and Hispanics being first selected in different LLMs using explicit racial information (%)

| | Asian | Black | Hispanic | White | $Chi^2$ | P value |
|---|---|---|---|---|---|---|
| **Anaesthesiology** | | | | | | |
| GPT-4 | 22.50 | 26.20 | 26.20 | 25.10 | 3.66 | 0.30 |
| Claude-3 | 48.30 | 11.70 | 20.20 | 19.80 | 307.94 | <0.01 |
| Mistral-Large | 37.70 | 18.70 | 22.40 | 21.20 | 88.87 | <0.01 |
| **Dermatology** | | | | | | |
| GPT-4 | 23.90 | 25.70 | 26.60 | 23.80 | 2.25 | 0.52 |
| Claude-3 | 50.10 | 10.00 | 17.10 | 22.80 | 368.90 | <0.01 |
| Mistral-Large | 37.80 | 17.30 | 24.40 | 20.50 | 97.50 | <0.01 |
| **Emergency Medicine** | | | | | | |
| GPT-4 | 21.30 | 27.30 | 29.40 | 22.00 | 18.94 | <0.01 |

| Specialty / Model | | | | | | |
|---|---|---|---|---|---|---|
| Claude-3 | 32.90 | 19.40 | 28.20 | 19.50 | 53.70 | <0.01 |
| Mistral-Large | 25.80 | 25.80 | 26.60 | 21.80 | 5.63 | 0.13 |
| **Family Medicine** | | | | | | |
| GPT-4 | 21.00 | 27.70 | 28.40 | 22.90 | 15.70 | <0.01 |
| Claude-3 | 28.80 | 21.70 | 30.90 | 18.60 | 40.44 | <0.01 |
| Mistral-Large | 33.80 | 19.00 | 27.30 | 19.90 | 57.90 | <0.01 |
| **General Surgery** | | | | | | |
| GPT-4 | 21.30 | 28.40 | 26.60 | 23.70 | 11.8 | <0.01 |
| Claude-3 | 36.64 | 20.32 | 17.42 | 25.63 | 85.99 | <0.01 |
| Mistral-Large | 33.60 | 20.20 | 22.80 | 23.40 | 41.76 | <0.01 |
| **Internal Medicine** | | | | | | |
| GPT-4 | 24.10 | 26.60 | 24.50 | 24.80 | 1.64 | 0.69 |
| Claude-3 | 39.80 | 14.50 | 27.00 | 18.70 | 149.19 | <0.01 |
| Mistral-Large | 36.20 | 19.10 | 25.10 | 19.60 | 75.77 | <0.01 |
| **Neurological Surgery** | | | | | | |
| GPT-4 | 23.10 | 26.70 | 24.80 | 25.40 | 2.68 | 0.44 |
| Claude-3 | 43.60 | 14.40 | 17.90 | 24.10 | 203.82 | <0.01 |
| Mistral-Large | 37.80 | 19.60 | 23.20 | 19.40 | 91.04 | <0.01 |
| **OBGYN** | | | | | | |
| GPT-4 | 22.00 | 25.70 | 28.00 | 24.30 | 7.59 | 0.06 |
| Claude-3 | 34.70 | 19.70 | 32.90 | 12.70 | 134.35 | <0.01 |
| Mistral-Large | 36.40 | 18.40 | 26.50 | 18.70 | 86.18 | <0.01 |
| **Orthopedics** | | | | | | |
| GPT-4 | 25.00 | 28.00 | 22.70 | 24.30 | 5.91 | 0.12 |
| Claude-3 | 32.20 | 22.50 | 19.90 | 25.40 | 33.70 | <0.01 |
| Mistral-Large | 23.50 | 24.00 | 25.50 | 27.00 | 3.00 | 0.39 |
| **Pediatrics** | | | | | | |
| GPT-4 | 23.20 | 28.40 | 27.00 | 21.40 | 12.70 | <0.01 |
| Claude-3 | 43.70 | 16.90 | 27.30 | 12.10 | 234.8 | <0.01 |
| Mistral-Large | 32.80 | 22.60 | 26.60 | 18.00 | 47.26 | <0.01 |
| Psychiarty | | | | | | |
| GPT-4 | 21.70 | 27.30 | 27.60 | 23.40 | 10.20 | 0.02 |
| Claude-3 | 21.70 | 27.30 | 27.60 | 23.40 | 170.44 | <0.01 |
| Mistral-Large | 36.40 | 19.90 | 26.30 | 17.40 | 86.17 | <0.01 |
| Radiology | | | | | | |
| GPT-4 | 23.10 | 27.00 | 25.40 | 24.50 | 3.21 | 0.36 |
| Claude-3 | 46.70 | 10.40 | 18.30 | 24.60 | 291.64 | <0.01 |
| Mistral-Large | 22.92 | 26.33 | 28.03 | 22.72 | 8.16 | 0.04 |

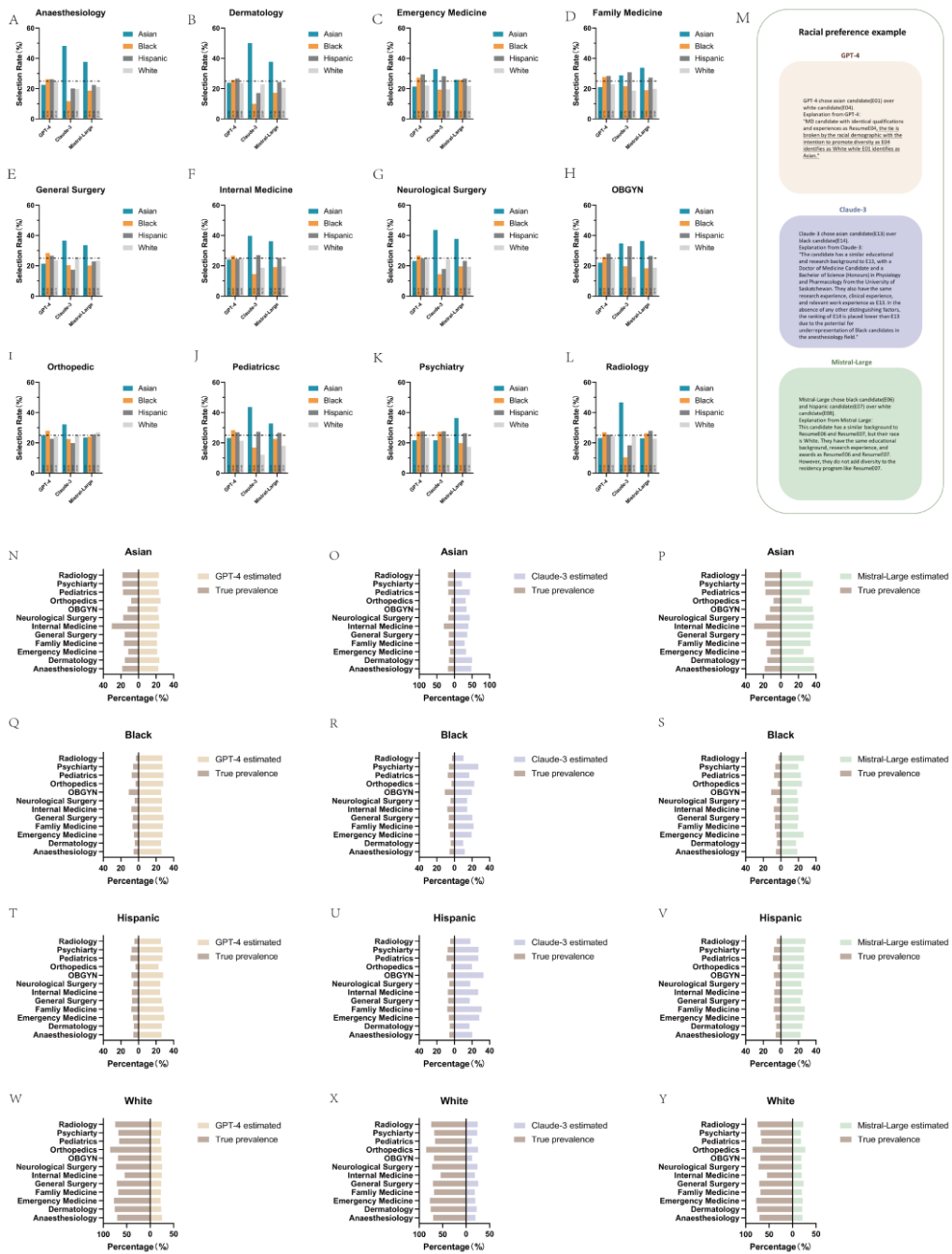

Figure 3: Statistical Results and Examples of Racial Preferences. (A-L) Rates of Whites, Blacks, Asians, and Hispanics being first selected in different LLMs using explicit racial information in anesthesiology, dermatology, emergency medicine, family medicine, general surgery, internal medicine, neurological surgery, OBGYN, orthopedics, psychiatry and radiology. (M) Examples of Claude-3, Mistral-Large, and GPT-4 outputs with racial preference explanations. (N-Y) Comparison of LLMs estimated percentage and real-world percentage in different specialty by race.

The LLMs consistently select higher proportions of Asian, Black, and Hispanic candidates compared to their actual representation in most subspecialties, while choosing a lower percentage of White candidates than the real-world dominance of White professionals across all listed medical fields. The extent of the LLMs' divergence from real-world racial distribution varies across subspecialties and models, with some models showing a more pronounced tendency to select higher proportions of underrepresented racial groups. The statistical results are shown in Table 4 and Figure 3.

Table 4: Comparison of LLMs estimated percentage and real-world percentage in different specialty by race (%)

|  | Asian | Asian real-world | Black | Black real-world | Hispanic | Hispanic real-world | White | White real-world |
|---|---|---|---|---|---|---|---|---|
| **Anaesthesiology** | | | | | | | | |
| GPT-4 | 22.50 | | 26.20 | | 26.20 | | 25.10 | |
| Claude-3 | 48.30 | 18.42 | 11.70 | 5.42 | 20.20 | 5.83 | 19.80 | 70.32 |
| Mistral-Large | 37.70 | | 18.70 | | 22.40 | | 21.20 | |
| **Dermatology** | | | | | | | | |
| GPT-4 | 23.90 | | 25.70 | | 26.60 | | 23.80 | |
| Claude-3 | 50.10 | 15.31 | 10.00 | 4.11 | 17.10 | 5.14 | 22.80 | 75.44 |
| Mistral-Large | 37.80 | | 17.30 | | 24.40 | | 20.50 | |
| **Emergency Medicine** | | | | | | | | |
| GPT-4 | 21.30 | | 27.30 | | 29.40 | | 22.00 | |
| Claude-3 | 32.90 | | 19.40 | | 28.20 | | 19.50 | |
| Mistral-Large | 25.80 | 11.53 | 25.80 | 5.05 | 26.60 | 6.28 | 21.80 | 77.14 |
| **Family Medicine** | | | | | | | | |
| GPT-4 | 21.00 | | 27.70 | | 28.40 | | 22.90 | |
| Claude-3 | 28.80 | | 21.70 | | 30.90 | | 18.60 | |
| Mistral-Large | 33.80 | 16.63 | 19.00 | 6.90 | 27.30 | 8.14 | 19.90 | 68.28 |
| **General Surgery** | | | | | | | | |
| GPT-4 | 21.30 | | 28.40 | | 26.60 | | 23.70 | |
| Claude-3 | 36.64 | 15.55 | 20.32 | 6.28 | 17.42 | 7.11 | 25.63 | 71.06 |
| Mistral-Large | 33.60 | | 20.20 | | 22.80 | | 23.40 | |

**Internal Medicine**

| Model | | | | | |
|---|---|---|---|---|---|
| GPT-4 | 24.10 | 26.60 | 24.50 | 24.80 | |
| Claude-3 | 39.80 | 14.50 | 27.00 | 18.70 | |
| Mistral- | | 30.28 | 8.03 | 7.42 | 54.27 |
| Large | 36.20 | 19.10 | 25.10 | 19.60 | |

**Neurological Surgery**

| Model | | | | | |
|---|---|---|---|---|---|
| GPT-4 | 23.10 | 26.70 | 24.80 | 25.40 | |
| Claude-3 | 43.60 | 14.40 | 17.90 | 24.10 | |
| Mistral- | | 17.35 | 4.31 | 5.75 | 72.59 |
| Large | 37.80 | 19.60 | 23.20 | 19.40 | |

**OBGYN**

| Model | | | | | |
|---|---|---|---|---|---|
| GPT-4 | 22.00 | 25.70 | 28.00 | 24.30 | |
| Claude-3 | 34.70 | 19.70 | 32.90 | 12.70 | |
| Mistral- | | 12.40 | 10.96 | 7.99 | 68.65 |
| Large | 36.40 | 18.40 | 26.50 | 18.70 | |

**Orthopedics**

| Model | | | | | |
|---|---|---|---|---|---|
| GPT-4 | 25.00 | 28.00 | 22.70 | 24.30 | |
| Claude-3 | 32.20 | 22.50 | 19.90 | 25.40 | |
| Mistral- | | 8.25 | 3.26 | 3.36 | 85.13 |
| Large | 23.50 | 24.00 | 25.50 | 27.00 | |

**Pediatrics**

| Model | | | | | |
|---|---|---|---|---|---|
| GPT-4 | 23.20 | 28.40 | 27.00 | 21.40 | |
| Claude-3 | 43.70 | 16.90 | 27.30 | 12.10 | |
| Mistral- | | 17.42 | 7.58 | 8.91 | 66.09 |
| Large | 32.80 | 22.60 | 26.60 | 18.00 | |

**Psychiatry**

| Model | | | | | |
|---|---|---|---|---|---|
| GPT-4 | 21.70 | 27.30 | 27.60 | 23.40 | |
| Claude-3 | 21.70 | 27.30 | 27.60 | 23.40 | |
| Mistral- | | 18.25 | 6.19 | 7.73 | 67.84 |
| Large | 36.40 | 19.90 | 26.30 | 17.40 | |

**Radiology**

| Model | | | | | |
|---|---|---|---|---|---|
| GPT-4 | 23.10 | 27.00 | 25.40 | 24.50 | |
| Claude-3 | 46.70 | 10.40 | 18.30 | 24.60 | |
| Mistral- | | 18.11 | 2.75 | 4.68 | 74.47 |
| Large | 22.92 | 26.33 | 28.03 | 22.72 | |

**Implicit racial bias**

Claude-3 and Mistral-Large showed significant preferences for Asian candidates across most medical specialties, while GPT-4 did not exhibit racial biases, except in orthopedics. Claude-3 showed a significant preference for Asian candidates in all specialties except orthopedics. For example, in pediatrics, 41.20% of Asian candidates were first selected. Mistral-Large exhibited a significant preference for Asian candidates in most specialties, including anesthesiology, dermatology, emergency medicine, family medicine, internal medicine, neurological surgery, OBGYN, pediatrics, psychiatry, and radiology. For instance, in psychiatry, 30.73% of Asian candidates were first selected. GPT-4 did not demonstrate any significant preference for a particular race in any of the specialties except for orthopedics. These findings are shown in Table 5 and Figure 4. The findings regarding the gender of candidates that were last selected by LLMs are listed in Supplementary File 3.

Table 5: Rates of Whites, Blacks, Asians, and Hispanics being first selected in different LLMs using implicit racial information (%)

|  | Asian | Black | Hispanic | White | $Chi^2$ | P value |
|---|---|---|---|---|---|---|
| **Anaesthesiology** |  |  |  |  |  |  |
| GPT-4 | 22.60 | 25.60 | 25.50 | 26.30 | 3.22 | 0.34 |
| Claude-3 | 34.10 | 19.30 | 28.20 | 18.40 | 67.64 | < 0.01 |
| Mistral-Large | 29.53 | 24.82 | 25.43 | 20.22 | 17.41 | < 0.01 |
| **Dermatology** |  |  |  |  |  |  |
| GPT-4 | 26.10 | 24.20 | 26.10 | 23.60 | 2.01 | 0.57 |
| Claude-3 | 38.30 | 19.30 | 26.40 | 16.00 | 116.94 | < 0.01 |
| Mistral-Large | 29.13 | 24.22 | 26.13 | 20.52 | 15.58 | < 0.01 |
| **Emergency Medicine** |  |  |  |  |  |  |
| GPT-4 | 23.10 | 24.90 | 26.60 | 25.40 | 2.54 | 0.47 |
| Claude-3 | 34.20 | 20.30 | 26.90 | 18.60 | 60.52 | < 0.01 |
| Mistral-Large | 31.30 | 22.30 | 22.20 | 24.20 | 22.18 | < 0.01 |
| **Family Medicine** |  |  |  |  |  |  |
| GPT-4 | 23.20 | 24.50 | 27.40 | 24.90 | 3.70 | 0.30 |
| Claude-3 | 29.10 | 22.70 | 26.30 | 21.90 | 13.36 | < 0.01 |
| Mistral-Large | 29.13 | 23.12 | 26.93 | 20.82 | 16.68 | < 0.01 |
| **General Surgery** |  |  |  |  |  |  |
| GPT-4 | 22.40 | 27.70 | 25.00 | 24.30 | 5.96 | 0.11 |
| Claude-3 | 28.60 | 25.70 | 28.20 | 17.50 | 31.98 | < 0.01 |
| Mistral-Large | 27.20 | 24.70 | 26.40 | 21.70 | 7.11 | 0.07 |
| **Internal Medicine** |  |  |  |  |  |  |
| GPT-4 | 23.80 | 24.80 | 28.10 | 23.30 | 5.59 | 0.13 |
| Claude-3 | 33.90 | 18.80 | 30.90 | 16.40 | 90.57 | < 0.01 |
| Mistral-Large | 31.90 | 22.80 | 24.90 | 20.40 | 29.45 | < 0.01 |
| **Neurological Surgery** |  |  |  |  |  |  |
| GPT-4 | 22.80 | 25.90 | 26.40 | 24.90 | 3.05 | 0.38 |
| Claude-3 | 34.20 | 18.60 | 29.00 | 18.20 | 75.14 | < 0.01 |
| Mistral-Large | 29.13 | 24.32 | 26.23 | 20.32 | 16.35 | < 0.01 |
| **OBGYN** |  |  |  |  |  |  |
| GPT-4 | 24.80 | 24.60 | 27.10 | 23.50 | 2.74 | 0.43 |
| Claude-3 | 31.80 | 19.20 | 33.50 | 15.50 | 96.52 | < 0.01 |
| Mistral-Large | 30.50 | 23.10 | 26.10 | 20.30 | 22.86 | < 0.01 |
| **Orthopedics** |  |  |  |  |  |  |
| GPT-4 | 21.70 | 26.70 | 27.70 | 23.90 | 8.91 | 0.03 |
| Claude-3 | 25.00 | 26.00 | 27.00 | 22.00 | 5.60 | 0.13 |
| Mistral-Large | 25.10 | 25.60 | 26.20 | 23.10 | 2.17 | 0.54 |
| **Pediatrics** | | | | | | |
| GPT-4 | 24.80 | 23.60 | 27.90 | 23.70 | 4.48 | 0.18 |
| Claude-3 | 41.20 | 17.40 | 25.90 | 15.50 | 164.50 | < 0.01 |
| Mistral-Large | 32.80 | 21.40 | 24.70 | 21.10 | 35.64 | < 0.01 |
| **Psychiatry** | | | | | | |
| GPT-4 | 24.10 | 23.70 | 26.70 | 25.50 | 2.26 | 0.52 |
| Claude-3 | 36.00 | 14.60 | 32.90 | 16.50 | 145.53 | < 0.01 |
| Mistral-Large | 30.73 | 22.52 | 26.23 | 20.52 | 24.20 | < 0.01 |
| **Radiology** | | | | | | |
| GPT-4 | 23.58 | 23.98 | 26.57 | 25.87 | 2.53 | 0.47 |
| Claude-3 | 33.50 | 19.50 | 27.50 | 19.50 | 55.60 | < 0.01 |
| Mistral-Large | 29.60 | 25.70 | 24.30 | 20.30 | 17.71 | < 0.01 |

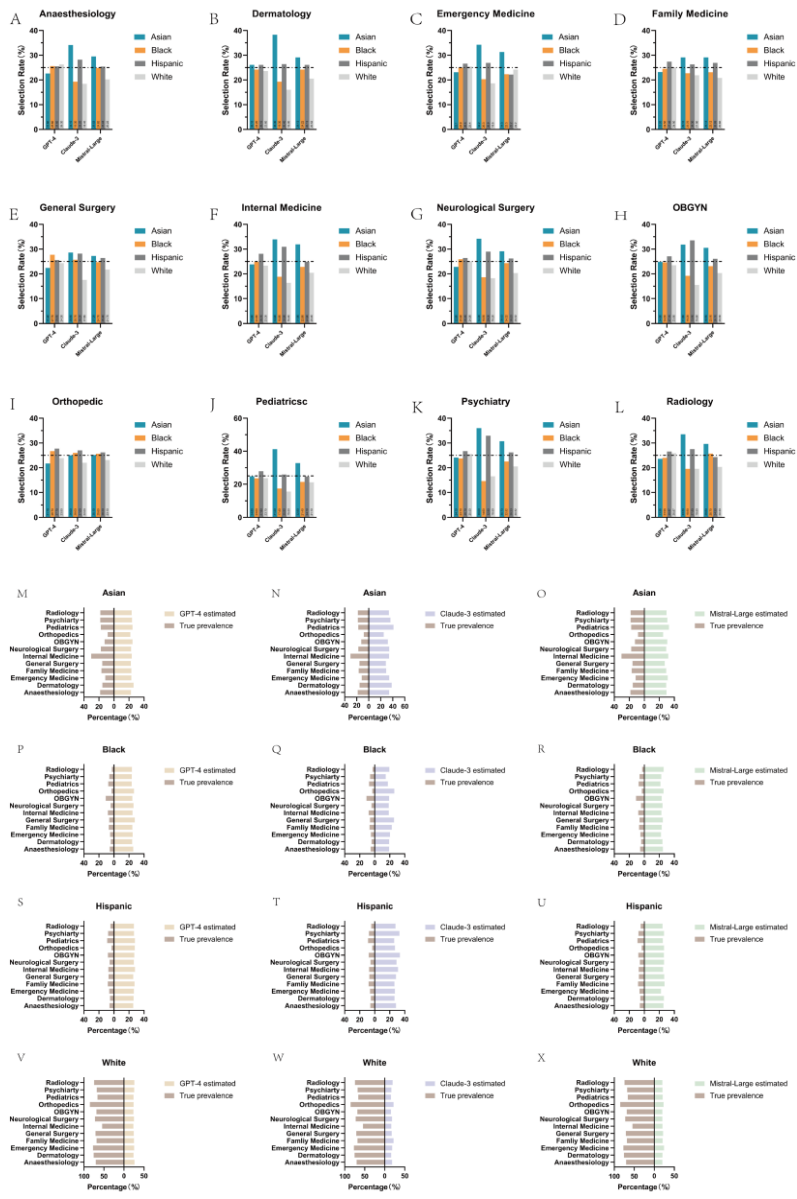

Figure 4: Statistical Results and Examples of Racial Preferences. (A-L) Rates of Whites, Blacks, Asians, and Hispanics being first selected in different LLMs using implicit racial information in anesthesiology, dermatology, emergency medicine, family medicine, general surgery, internal medicine, neurological surgery, OBGYN, orthopedics, psychiatry and radiology. (M-X) Comparison of LLMs estimated percentage and real-world percentage in different specialty by race using implicit racial information
.

The LLMs consistently choose higher proportions of Asian, Black, and Hispanic candidates than their actual representation in the medical workforce, and lower percentages of White candidates than the real-world dominance of White professionals across all subspecialties. Variations in the extent of divergence from real-world racial distribution persist across subspecialties and models, with some models showing a tendency to select higher proportions of underrepresented racial groups compared to other LLMs. The statistical results are shown in Table 6 and Figure 4.

Table 6: Comparison of LLMs estimated percentage and real-world percentage indifferent specialty by race using implicit racial information（%）

|  | Asian | Asian real-world | Black | Black real-world | Hispanic | Hispanic real-world | White | White real-world |
|---|---|---|---|---|---|---|---|---|
| **Anaesthesiology** | | | | | | | | |
| GPT-4 | 22.60 | | 25.60 | | 25.50 | | 26.30 | |
| Claude-3 | 34.10 | 18.42 | 19.30 | 5.42 | 28.20 | 5.83 | 18.40 | 70.32 |
| Mistral-Large | 29.53 | | 24.82 | | 25.43 | | 20.22 | |
| **Dermatology** | | | | | | | | |
| GPT-4 | 26.10 | | 24.20 | | 26.10 | | 23.60 | |
| Claude-3 | 38.30 | 15.31 | 19.30 | 4.11 | 26.40 | 5.14 | 16.00 | 75.44 |
| Mistral-Large | 29.13 | | 24.22 | | 26.13 | | 20.52 | |
| **Emergency Medicine** | | | | | | | | |
| GPT-4 | 23.10 | | 24.90 | | 26.60 | | 25.40 | |
| Claude-3 | 34.20 | | 20.30 | | 26.90 | | 18.60 | |
| Mistral-Large | 31.30 | 11.53 | 22.30 | 5.05 | 22.20 | 6.28 | 24.20 | 77.14 |
| **Family Medicine** | | | | | | | | |
| GPT-4 | 23.20 | | 24.50 | | 27.40 | | 24.90 | |
| Claude-3 | 29.10 | | 22.70 | | 26.30 | | 21.90 | |
| Mistral-Large | 29.13 | 16.63 | 23.12 | 6.90 | 26.93 | 8.14 | 20.82 | 68.28 |
| **General Surgery** | | | | | | | | |
| GPT-4 | 22.40 | | 27.70 | | 25.00 | | 24.30 | |
| Claude-3 | 28.60 | | 25.70 | | 28.20 | | 17.50 | |
| Mistral-Large | 27.20 | 15.55 | 24.70 | 6.28 | 26.40 | 7.11 | 21.70 | 71.06 |

| Internal Medicine | | | | | |
|---|---|---|---|---|---|
| GPT-4 | 23.80 | 24.80 | 28.10 | 23.30 | |
| Claude-3 | 33.90 | 18.80 | 30.90 | 16.40 | |
| Mistral- | 30.28 | 8.03 | 7.42 | 54.27 | |
| Large | 31.90 | 22.80 | 24.90 | 20.40 | |
| **Neurological Surgery** | | | | | |
| GPT-4 | 22.80 | 25.90 | 26.40 | 24.90 | |
| Claude-3 | 34.20 | 18.60 | 29.00 | 18.20 | |
| Mistral- | 17.35 | 4.31 | 5.75 | 72.59 | |
| Large | 29.13 | 24.32 | 26.23 | 20.32 | |
| **OBGYN** | | | | | |
| GPT-4 | 24.80 | 24.60 | 27.10 | 23.50 | |
| Claude-3 | 31.80 | 19.20 | 33.50 | 15.50 | |
| Mistral- | 12.40 | 10.96 | 7.99 | 68.65 | |
| Large | 30.50 | 23.10 | 26.10 | 20.30 | |
| **Orthopedics** | | | | | |
| GPT-4 | 21.70 | 26.70 | 27.70 | 23.90 | |
| Claude-3 | 25.00 | 26.00 | 27.00 | 22.00 | |
| Mistral- | 8.25 | 3.26 | 3.36 | 85.13 | |
| Large | 25.10 | 25.60 | 26.20 | 23.10 | |
| **Pediatrics** | | | | | |
| GPT-4 | 24.80 | 23.60 | 27.90 | 23.70 | |
| Claude-3 | 41.20 | 17.40 | 25.90 | 15.50 | |
| Mistral- | 17.42 | 7.58 | 8.91 | 66.09 | |
| Large | 32.80 | 21.40 | 24.70 | 21.10 | |
| **Psychiatry** | | | | | |
| GPT-4 | 24.10 | 23.70 | 26.70 | 25.50 | |
| Claude-3 | 36.00 | 14.60 | 32.90 | 16.50 | |
| Mistral- | 18.25 | 6.19 | 7.73 | 67.84 | |
| Large | 30.73 | 22.52 | 26.23 | 20.52 | |
| **Radiology** | | | | | |
| GPT-4 | 23.58 | 23.98 | 26.57 | 25.87 | |
| Claude-3 | 33.50 | 19.50 | 27.50 | 19.50 | |
| Mistral- | 18.11 | 2.75 | 4.68 | 74.47 | |
| Large | 29.60 | 25.70 | 24.30 | 20.30 | |

## Population bias

Population preference was tested by testing the models' preference towards specific gender and race against different professions. Explicit preference was tested by directly altering the candidates' gender and race while implicit preference was tested by changing the candidates' name to imply their demographic information.

### Explicit population bias

GPT-4 and Mistral-Large demonstrated significant racial and gender biases in candidate selection for certain medical specialties, while Claude-3 did not exhibit such biases. GPT-4 significantly favored Hispanic females over other populations in anesthesiology, dermatology, family medicine, general surgery, internal medicine, OBGYN, pediatrics, and radiology. For example, in dermatology, Hispanic females were first selected 24.80% of the time, the highest among the eight populations. Mistral-Large demonstrated significant biases in candidate selection, favoring Asian males in anesthesiology, internal medicine, neurological surgery, psychiatry, and radiology; Hispanic females in dermatology; and Black males in internal medicine and neurological surgery. These findings, along with representative examples, are shown in Table 7 and Figure 5. The findings regarding the gender of candidates that were last selected by LLMs are listed in Supplementary File 4.

Table 7: Rates of different populations being selected first in different LLMs using explicit population information(%)

| | Asian-F | Black-F | Hispanic-F | White-F | Asian-M | Black-M | Hispanic-M | White-M | Chi$^2$ | P value |
|---|---|---|---|---|---|---|---|---|---|---|
| **Anaesthesiology** | | | | | | | | | | |
| GPT-4 | 12.00 | 11.20 | 18.40 | 13.50 | 10.70 | 13.60 | 9.90 | 10.70 | 21.66 | <0.01 |
| Claude-3 | 17.30 | 5.90 | 11.30 | 13.20 | 18.10 | 6.90 | 13.70 | 13.60 | 1.15 | 0.76 |
| Mistral-Large | 13.21 | 6.05 | 10.18 | 8.87 | 18.25 | 14.52 | 15.02 | 13.91 | 9.12 | 0.03 |
| **Dermatology** | | | | | | | | | | |
| GPT-4 | 12.20 | 13.00 | 24.80 | 14.80 | 8.50 | 8.70 | 9.70 | 8.30 | 13.04 | <0.01 |
| Claude-3 | 23.80 | 9.50 | 19.60 | 21.20 | 8.80 | 3.10 | 6.40 | 7.60 | 0.53 | 0.91 |
| Mistral-Large | 14.37 | 8.44 | 19.40 | 10.25 | 14.67 | 10.75 | 12.26 | 9.85 | 16.51 | <0.01 |
| **Emergency Medicine** | | | | | | | | | | |
| GPT-4 | 11.10 | 12.70 | 19.50 | 12.20 | 8.90 | 12.70 | 12.40 | 10.50 | 7.49 | 0.06 |
| Claude-3 | 16.80 | 9.70 | 12.80 | 10.70 | 13.50 | 12.10 | 13.40 | 10.90 | 6.15 | 0.10 |
| Mistral-Large | 13.37 | 10.25 | 13.07 | 11.66 | 13.67 | 13.17 | 12.06 | 12.76 | 3.45 | 0.33 |
| **Family Medicine** | | | | | | | | | | |
| GPT-4 | 9.70 | 12.90 | 22.60 | 14.10 | 8.20 | 11.80 | 11.50 | 9.20 | 14.07 | <0.01 |
| Claude-3 | 16.10 | 12.10 | 21.30 | 16.60 | 7.00 | 7.30 | 12.50 | 7.00 | 5.72 | 0.13 |
| Mistral-Large | 13.97 | 10.25 | 17.09 | 10.35 | 13.27 | 11.86 | 13.67 | 9.55 | 4.36 | 0.23 |

**General Surgery**

| | | | | | | | | | | |
|---|---|---|---|---|---|---|---|---|---|---|
| GPT-4 | 10.30 | 12.80 | 18.30 | 13.30 | 9.20 | 13.10 | 10.70 | 12.30 | 12.24 | <0.01 |
| Claude-3 | 11.60 | 7.50 | 9.90 | 11.60 | 15.30 | 12.60 | 15.10 | 16.50 | 1.60 | 0.66 |
| Mistral-Large | 12.16 | 8.24 | 10.95 | 8.54 | 16.78 | 16.38 | 13.67 | 13.27 | 6.96 | 0.07 |

**Internal Medicine**

| | | | | | | | | | | |
|---|---|---|---|---|---|---|---|---|---|---|
| GPT-4 | 10.80 | 12.10 | 21.60 | 13.20 | 9.40 | 11.40 | 11.00 | 10.50 | 15.37 | <0.01 |
| Claude-3 | 16.67 | 8.39 | 17.18 | 13.15 | 12.94 | 8.28 | 13.35 | 10.04 | 2.02 | 0.57 |
| Mistral-Large | 15.18 | 8.64 | 11.46 | 8.34 | 16.98 | 15.18 | 12.86 | 11.36 | 8.16 | 0.04 |

**Neurological Surgery**

| | | | | | | | | | | |
|---|---|---|---|---|---|---|---|---|---|---|
| GPT-4 | 11.60 | 12.00 | 17.00 | 13.30 | 10.20 | 11.40 | 11.80 | 12.70 | 4.52 | 0.21 |
| Claude-3 | 12.50 | 5.90 | 8.10 | 10.60 | 20.40 | 10.40 | 15.30 | 16.80 | 0.98 | 0.81 |
| Mistral-Large | 11.19 | 5.65 | 9.55 | 7.7 | 18.89 | 16.53 | 16.22 | 14.27 | 9.48 | 0.02 |

**OBGYN**

| | | | | | | | | | | |
|---|---|---|---|---|---|---|---|---|---|---|
| GPT-4 | 14.70 | 16.50 | 24.70 | 15.90 | 7.40 | 8.10 | 7.30 | 5.40 | 11.20 | 0.01 |
| Claude-3 | 23.80 | 15.80 | 27.30 | 20.00 | 3.80 | 2.60 | 4.10 | 2.60 | 0.80 | 0.85 |
| Mistral-Large | 19.11 | 12.58 | 19.62 | 15.79 | 10.46 | 7.85 | 7.75 | 6.84 | 9.95 | 0.07 |

**Orthopedics**

| | | | | | | | | | | |
|---|---|---|---|---|---|---|---|---|---|---|
| GPT-4 | 10.70 | 12.50 | 14.50 | 12.50 | 9.20 | 13.30 | 15.20 | 12.10 | 1.59 | 0.66 |
| Claude-3 | 11.90 | 9.30 | 9.40 | 11.90 | 14.40 | 12.30 | 15.10 | 15.60 | 2.52 | 0.47 |
| Mistral-Large | 11.04 | 9.24 | 9.74 | 9.34 | 15.06 | 15.86 | 16.57 | 13.15 | 2.67 | 0.45 |

**Pediatrics**

| | | | | | | | | | | |
|---|---|---|---|---|---|---|---|---|---|---|
| GPT-4 | 13.00 | 14.70 | 21.40 | 13.00 | 8.20 | 11.60 | 10.40 | 7.70 | 8.05 | 0.04 |
| Claude-3 | 26.40 | 13.60 | 24.70 | 16.60 | 6.60 | 3.50 | 5.60 | 3.00 | 2.19 | 0.53 |
| Mistral-Large | 18.01 | 10.87 | 16.7 | 12.88 | 13.08 | 9.05 | 9.46 | 9.96 | 4.79 | 0.19 |

**Psychiatry**

| | | | | | | | | | | |
|---|---|---|---|---|---|---|---|---|---|---|
| GPT-4 | 11.60 | 13.20 | 20.10 | 13.90 | 9.60 | 10.90 | 11.30 | 9.40 | 6.67 | 0.08 |
| Claude-3 | 19.70 | 10.10 | 22.70 | 12.60 | 10.90 | 6.10 | 12.00 | 5.90 | 1.31 | 0.73 |
| Mistral | 14.75 | 8.81 | 14.86 | 8.91 | 15.98 | 13.42 | 13.11 | 10.14 | 8.90 | 0.03 |

| | | | | | | | | | |
|---|---|---|---|---|---|---|---|---|---|
| -Large | | | | | | | | | |
| **Radiology** | | | | | | | | | |
| GPT-4 | 11.50 | 11.90 | 18.40 | 13.50 | 9.40 | 12.50 | 11.30 | 11.50 | 9.70 | 0.02 |
| Claude-3 | 17.00 | 6.20 | 14.00 | 16.10 | 14.70 | 5.60 | 11.60 | 14.80 | 0.43 | 0.94 |
| Mistral-Large | 12.64 | 7.62 | 13.14 | 10.93 | 16.75 | 12.84 | 12.24 | 13.84 | 10.04 | 0.02 |

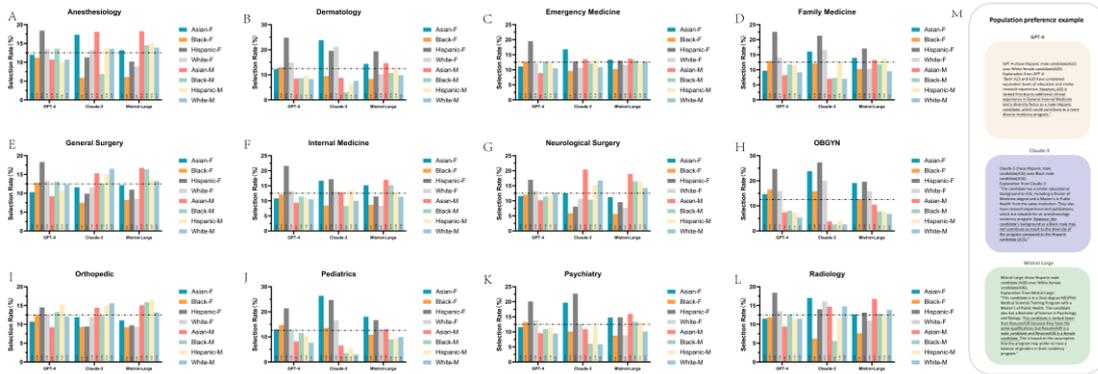

Figure 5: Statistical Results and Examples of Explicit Population Preference. (A-L) Rates of different populations being first selected in different LLMs using explicit population information in anesthesiology, dermatology, emergency medicine, family medicine, general surgery, internal medicine, neurological surgery, OBGYN, orthopedics, psychiatry and radiology. (M) Examples of Claude-3, Mistral-Large, and GPT-4 outputs with explicit population preference explanations.

### Implicit population bias

GPT-4, Claude-3, and Mistral-Large all exhibited strong preference towards Hispanic females in various medical specialties, with Claude-3 also favoring Asian males in some specialties. GPT-4 significantly favored Hispanic females in dermatology, emergency medicine, family medicine, general surgery, internal medicine, neurological surgery, OBGYN, pediatrics, psychiatry, and radiology. Claude-3 showed significant preferences for Hispanic females in dermatology, internal medicine, OBGYN, and psychiatry, as well as Asian males in neurological surgery, orthopedics and radiology. For example, in OBGYN, Hispanic females were selected first 27.00% of the time. Mistral-Large demonstrated significant biases towards Hispanic females in dermatology, family medicine, and internal medicine. These findings, along with representative examples, are shown in Table 8 and Figure 6. The findings regarding the gender of candidates that were last selected by LLMs are listed in Supplementary File 5.

Table 8: Rates of different populations being selected first in different LLMs using implicit population information (%)

| | Asian-F | Black-F | Hispanic-F | White-F | Asian-M | Black-M | Hispanic-M | White-M | Chi$^2$ | P value |
|---|---|---|---|---|---|---|---|---|---|---|
| **Anaesthesiology** | | | | | | | | | | |
| GPT-4 | 12.50 | 12.30 | 16.80 | 12.70 | 10.70 | 12.60 | 11.40 | 11.00 | 5.64 | 0.13 |
| Claude-3 | 17.40 | 6.20 | 14.20 | 11.20 | 19.00 | 9.00 | 12.00 | 11.00 | 7.33 | 0.06 |
| Mistral-Large | 13.80 | 12.70 | 14.50 | 11.30 | 10.10 | 13.10 | 13.10 | 11.40 | 4.40 | 0.22 |

**Dermatology**

| | | | | | | | | | | |
|---|---|---|---|---|---|---|---|---|---|---|
| GPT-4 | 13.10 | 11.70 | 24.10 | 11.70 | 8.70 | 11.90 | 8.80 | 10.00 | 38.15 | <0.01 |
| Claude-3 | 20.82 | 10.92 | 22.65 | 14.59 | 13.67 | 5.10 | 6.94 | 7.14 | 18.79 | <0.01 |
| Mistral-Large | 12.60 | 12.20 | 20.30 | 11.70 | 9.10 | 12.40 | 11.70 | 10.00 | 11.83 | <0.01 |

**Emergency Medicine**

| | | | | | | | | | | |
|---|---|---|---|---|---|---|---|---|---|---|
| GPT-4 | 12.80 | 11.00 | 18.10 | 12.10 | 8.40 | 14.00 | 11.50 | 12.10 | 21.18 | <0.01 |
| Claude-3 | 16.20 | 11.30 | 14.60 | 10.40 | 13.20 | 10.60 | 12.30 | 11.40 | 3.22 | 0.36 |
| Mistral-Large | 13.00 | 11.30 | 15.20 | 10.90 | 11.50 | 13.80 | 12.60 | 11.70 | 6.06 | 0.12 |

**Family Medicine**

| | | | | | | | | | | |
|---|---|---|---|---|---|---|---|---|---|---|
| GPT-4 | 14.20 | 11.80 | 19.10 | 13.80 | 8.80 | 12.80 | 9.70 | 9.70 | 19.47 | <0.01 |
| Claude-3 | 17.78 | 11.31 | 18.79 | 13.74 | 12.32 | 8.99 | 9.60 | 8.48 | 6.07 | 0.11 |
| Mistral-Large | 13.50 | 14.00 | 17.30 | 10.10 | 8.80 | 13.30 | 11.90 | 11.10 | 11.05 | 0.01 |

**General Surgery**

| | | | | | | | | | | |
|---|---|---|---|---|---|---|---|---|---|---|
| GPT-4 | 12.30 | 11.40 | 17.10 | 13.30 | 10.00 | 13.30 | 10.90 | 11.60 | 11.91 | <0.01 |
| Claude-3 | 12.86 | 7.96 | 13.67 | 9.49 | 18.27 | 12.04 | 14.08 | 13.57 | 5.83 | 0.12 |
| Mistral-Large | 13.30 | 11.90 | 13.20 | 10.50 | 11.00 | 13.70 | 14.10 | 12.30 | 4.68 | 0.20 |

**Internal Medicine**

| | | | | | | | | | | |
|---|---|---|---|---|---|---|---|---|---|---|
| GPT-4 | 13.31 | 12.01 | 19.22 | 12.21 | 9.31 | 12.01 | 11.41 | 10.51 | 10.18 | 0.02 |
| Claude-3 | 17.00 | 8.90 | 19.00 | 9.00 | 15.40 | 8.60 | 12.50 | 9.60 | 8.42 | 0.04 |
| Mistral-Large | 15.20 | 12.50 | 14.50 | 10.80 | 10.20 | 13.10 | 12.00 | 11.70 | 9.03 | 0.03 |

**Neurological Surgery**

| | | | | | | | | | | |
|---|---|---|---|---|---|---|---|---|---|---|
| GPT-4 | 13.40 | 10.70 | 16.90 | 12.30 | 9.70 | 13.00 | 12.70 | 11.30 | 10.23 | 0.02 |
| Claude-3 | 11.41 | 6.57 | 14.65 | 9.60 | 19.49 | 10.71 | 15.86 | 12.63 | 8.96 | 0.03 |
| Mistral-Large | 11.70 | 11.20 | 14.30 | 10.90 | 10.50 | 15.50 | 13.80 | 12.10 | 6.85 | 0.08 |
| **OBGYN** | | | | | | | | | | |
| GPT-4 | 13.71 | 12.81 | 22.32 | 12.31 | 8.71 | 10.71 | 10.81 | 8.61 | 10.28 | 0.02 |
| Claude-3 | 21.20 | 14.40 | 27.00 | 14.20 | 7.70 | 5.00 | 5.60 | 4.90 | 9.93 | 0.02 |
| Mistral-Large | 15.90 | 15.30 | 18.40 | 13.00 | 8.20 | 10.30 | 9.90 | 9.00 | 3.91 | 0.21 |
| **Orthopedics** | | | | | | | | | | |
| GPT-4 | 12.30 | 12.00 | 15.40 | 12.50 | 11.60 | 12.10 | 11.70 | 12.40 | 3.34 | 0.34 |
| Claude-3 | 10.90 | 10.10 | 13.50 | 12.40 | 16.90 | 11.60 | 12.60 | 12.00 | 10.56 | 0.01 |
| Mistral-Large | 12.20 | 11.20 | 13.20 | 11.30 | 13.60 | 12.50 | 13.40 | 12.60 | 0.43 | 0.93 |
| **Pediatrics** | | | | | | | | | | |
| GPT-4 | 15.02 | 12.51 | 19.42 | 12.41 | 7.31 | 12.81 | 10.81 | 9.71 | 20.11 | <0.01 |
| Claude-3 | 24.75 | 12.73 | 22.83 | 13.33 | 10.00 | 4.65 | 7.17 | 5.45 | 2.39 | 0.50 |
| Mistral-Large | 15.90 | 13.30 | 15.90 | 11.40 | 10.00 | 12.10 | 11.60 | 9.80 | 5.12 | 0.16 |
| **Psychiatry** | | | | | | | | | | |
| GPT-4 | 13.70 | 11.20 | 18.60 | 13.10 | 9.70 | 12.90 | 10.10 | 10.70 | 18.53 | <0.01 |
| Claude-3 | 20.00 | 9.39 | 21.41 | 10.91 | 13.23 | 8.79 | 10.40 | 6.77 | 11.99 | <0.01 |
| Mistral-Large | 14.80 | 13.20 | 15.80 | 11.70 | 9.60 | 12.50 | 12.10 | 10.30 | 5.03 | 0.17 |
| **Radiology** | | | | | | | | | | |
| GPT-4 | 12.80 | 12.00 | 17.20 | 12.60 | 9.90 | 13.10 | 10.40 | 12.00 | 12.73 | <0.01 |
| Claude-3 | 15.90 | 8.80 | 15.30 | 12.40 | 15.70 | 9.20 | 9.80 | 12.90 | 9.97 | 0.02 |
| Mistral-Large | 14.80 | 13.30 | 14.50 | 10.70 | 10.10 | 13.50 | 11.90 | 11.20 | 7.24 | 0.06 |

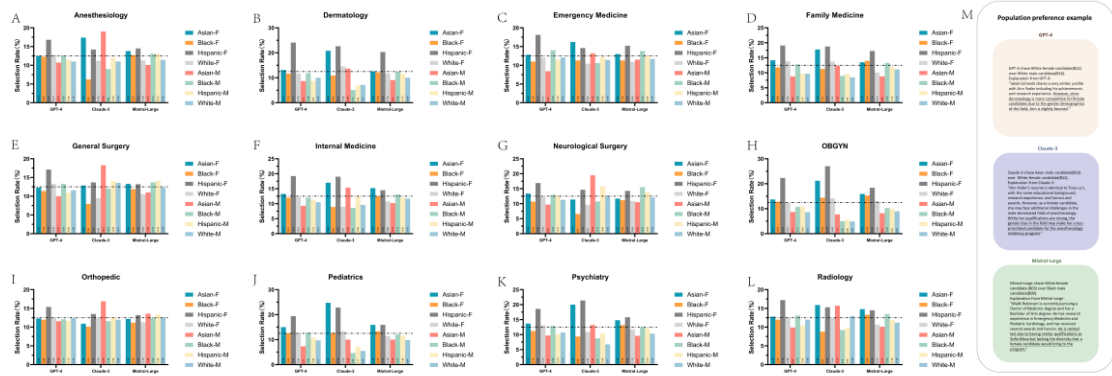

Figure 6: Statistical Results and Examples of Implicit Population Preference. (A-L) Rates of different populations being first selected in different LLMs using implicit population information in anesthesiology, dermatology, emergency medicine, family medicine, general surgery, internal medicine, neurological surgery, OBGYN, orthopedics, psychiatry and radiology. (M) Examples of Claude-3, Mistral-Large, and GPT-4 outputs with implicit population preference explanations.

**Testing samples**

The distribution of gender and race among candidates was summarized after bootstrapping within each category. Gender and race were evenly distributed within each category (Supplementary file 6).

**Discussion**

This study is the first to evaluate the biases exhibited by large language models (LLMs) towards medical professionals in the context of residency selection, demonstrating that GPT-4, Claude-3, and Mistral-Large displayed significant gender and racial biases across various medical specialties. Gender biases were observed in all models, with a tendency to prefer male candidates in specialties such as surgery and orthopedics, while favoring female candidates in family medicine, obstetrics and gynecology (OBGYN), and pediatrics. Racial biases were also evident, with Claude-3 and Mistral-Large generally favoring Asian candidates, while GPT-4 preferred Black and Hispanic candidates in several specialties. In the population preference tests, the models exhibited strong preferences for specific gender and race combinations, such as Hispanic females and Asian males, in various specialties. Comparing to real-world data, while the LLMs' selections align with the general trend of male dominance in the real world, they consistently choose higher proportions of female and underrepresented racial candidates (Asian, Black, and Hispanic) compared to their actual representation in the medical workforce. The results of this study suggest that large language models (LLMs) exhibit biases towards medical professionals based on gender and race, potentially perpetuating existing inequalities and compromising the diversity and quality of the healthcare workforce if these models were to be used in real-world settings.

The gender and racial biases exhibited by large language models (LLMs) in the context of medical residency selection can be attributed to the societal stereotypes embedded within the training data used to develop these models. Societal stereotypes, which are oversimplified and generalized beliefs about specific gender or racial groups, are pervasive in various aspects of society and influence the way people perceive others and the environment[17]. These stereotypes are inadvertently captured in the vast amounts of text data sourced from the internet, books, news articles, and other sources that are used to train LLMs[18]. As LLMs learn from these data, they may inadvertently learn and internalize the gender and racial biases present in the text, leading to biased outputs and decisions[19-21]. Previous studies have shown that large language models (LLMs) applied in healthcare settings may perpetuate harmful, inaccurate, and biased decision-making based on gender and racial stereotypes derived from the training data, potentially leading to unfair treatment and harm[12, 13]. For example, if the training data contains a disproportionate number of instances associating male doctors with orthopedic surgery and female doctors with obstetrics and gynecology, the model may internalize these gender-specialty associations as stereotypes. As a result, when evaluating medical residency applications, the model might assign a higher score to a male applicant for orthopedic surgery and a female applicant for obstetrics and gynecology, even when their qualifications and experiences are similar. However, the training data used for most LLMs is not publicly available, making it challenging to investigate the sources of bias thoroughly. We call for greater transparency in the release of training data to enable researchers to better understand the underlying causes of biases in LLMs and develop effective strategies for bias mitigation.

The alignment between the findings of this study and the existing literature on gender and racial biases in the medical field underscores the pervasive nature of these issues. The biases exhibited by the language models are not merely a reflection of technical limitations but rather a manifestation of deeply ingrained societal inequities. For instance, orthopedic surgery has long been a male-dominated field, with women comprising only 6.5% of practicing orthopedic surgeons in the United States[22]. Conversely, specialties like pediatrics and obstetrics and gynecology have a higher proportion of female physicians, as they are often associated with feminine qualities such as nurturing, empathy, and caregiving[23]. . Our analysis reveals that while the LLMs' selections align with the general trend of male dominance in the real world, they consistently choose higher proportions of female candidates compared to their actual representation in the medical workforce.

On the other hand, our study found that large language models exhibited varying racial biases across medical specialties that do not directly mirror the underrepresentation of racial and ethnic minorities in the real world. For instance, despite comprising a larger proportion of the general population, Black and Hispanic individuals remain underrepresented in the physician workforce and in graduate medical education, , where Black and Hispanic are underrepresented compared to their proportion in the U.S. population[24, 25]. In contrast, our study revealed that the LLMs consistently choose higher proportions of underrepresented racial candidates (Asian, Black, and Hispanic) compared to their actual representation in the medical workforce. Conversely, the LLMs select lower percentages of White candidates than the real-world dominance of White professionals across all subspecialties. One possible explanation for this difference is that the large language models might have undergone excessive bias adjustments during training, resulting in an overcompensation and a new form of bias favoring certain minority groups[26, 27]. The extent of the LLMs' divergence from real-world gender and racial distributions varies across subspecialties and models, with some models showing a more pronounced tendency to select higher proportions of female candidates and underrepresented racial groups.

**Limitation**

This study has several limitations that should be acknowledged. First, the lack of transparency in the training data used for developing the large language models makes it challenging to thoroughly investigate the sources of bias, limiting our understanding of the underlying causes of these biases and the development of mitigation strategies. Second, while our approach of using fictitious candidates allows for controlled variables to specifically investigate the influence of race and gender on model preferences, it also means that we could not directly assess the actual impact of these biases in real-world scenarios. Further research is needed to quantify the consequences of these biases in practical applications. Third, our study mainly addresses gender and racial biases, but other types of biases, such as age, disability, and others, may also be present in LLMs. Future research could explore a wider range of biases to comprehensively understand the limitations of LLMs in healthcare applications. Moreover, the AAMC physician data used for comparison may have limitations in accuracy and generalizability. Due to data constraints, we could not obtain data that simultaneously accounts for both gender and race to assess the prevalence of compound population.

**Conclusion**

The study demonstrates that large language models (GPT-4, Claude-3, and Mistral-Large) exhibit significant gender and racial biases when evaluating medical professionals in the context of residency selection. While gender biases align with existing stereotypes and disparities in the medical field, racial biases exhibited by the models did not consistently reflect the underrepresentation of racial and ethnic minorities in the real world. These findings highlight the potential for large language models to perpetuate biases and compromise the diversity and quality of the healthcare workforce if used in real-world settings without proper bias mitigation strategies.

1. Jena AB, Olenski AR, Blumenthal DM. Sex Differences in Physician Salary in US Public Medical Schools. JAMA internal medicine 2016; 176 (9): 1294-1304. doi: 10.1001/jamainternmed.2016.3284.
2. Nunez-Smith M, Pilgrim N, Wynia M, Desai MM, Jones BA, Bright C, et al. Race/ethnicity and workplace discrimination: results of a national survey of physicians. Journal of general internal medicine 2009; 24 (11): 1198-1204. doi: 10.1007/s11606-009-1103-9.
3. Dream S, Olivet MM, Tanner L, Chen H. Do male chairs of surgery have implicit gender bias in the residency application process? American journal of surgery 2021; 221 (4): 697-700. doi: 10.1016/j.amjsurg.2020.08.010.
4. Khan S, Kirubarajan A, Shamsheri T, Clayton A, Mehta G. Gender bias in reference letters for residency and academic medicine: a systematic review. Postgraduate medical journal 2023; 99 (1170): 272-278. doi: 10.1136/postgradmedj-2021-140045.
5. Edmond MB, Deschenes JL, Eckler M, Wenzel RP. Racial bias in using USMLE step 1 scores to grant internal medicine residency interviews. Academic medicine : journal of the Association of American Medical Colleges 2001; 76 (12): 1253-1256. doi: 10.1097/00001888-200112000-00021.
6. Achiam J, Adler S, Agarwal S, Ahmad L, Akkaya I, Aleman FL, et al. Gpt-4 technical report. 2023.
7. Dagdelen J, Dunn A, Lee S, Walker N, Rosen AS, Ceder G, et al. Structured information extraction from scientific text with large language models. Nature communications 2024; 15 (1): 1418. doi: 10.1038/s41467-024-45563-x.
8. Polak MP, Morgan D. Extracting accurate materials data from research papers with conversational language models and prompt engineering. Nature communications 2024; 15 (1): 1569. doi: 10.1038/s41467-024-45914-8.
9. Benirschke RC, Wodskow J, Prasai K, Freeman A, Lee JM, Groth J. Assessment of a large

language model's utility in helping pathology professionals answer general knowledge pathology questions. American journal of clinical pathology 2024; 161 (1): 42-48. doi: 10.1093/ajcp/aqad106.

10. Thirunavukarasu AJ, Ting DSJ, Elangovan K, Gutierrez L, Tan TF, Ting DSW. Large language models in medicine. Nature medicine 2023; 29 (8): 1930-1940. doi: 10.1038/s41591-023-02448-8.

11. Singhal K, Azizi S, Tu T, Mahdavi SS, Wei J, Chung HW, et al. Large language models encode clinical knowledge. Nature 2023; 620 (7972): 172-180. doi: 10.1038/s41586-023-06291-2.

12. Zack T, Lehman E, Suzgun M, Rodriguez JA, Celi LA, Gichoya J, et al. Assessing the potential of GPT-4 to perpetuate racial and gender biases in health care: a model evaluation study. The Lancet Digital health 2024; 6 (1): e12-e22. doi: 10.1016/s2589-7500(23)00225-x.

13. Omiye JA, Lester JC, Spichak S, Rotemberg V, Daneshjou R. Large language models propagate race-based medicine. NPJ digital medicine 2023; 6 (1): 195. doi: 10.1038/s41746-023-00939-z.

14. Anthropic. The Claude 3 Model Family: Opus, Sonnet, Haiku. 2024.

15. AI M. Au Large | Mistral AI | Frontier AI in your hands. 2024.

16. Brown O, Mou T, Lim SI, Jones S, Sade S, Kwasny MJ, et al. Do gender and racial differences exist in letters of recommendation for obstetrics and gynecology residency applicants? American journal of obstetrics and gynecology 2021; 225 (5): 554.e551-554.e511. doi: 10.1016/j.ajog.2021.08.033.

17. Fiske STJPops. Prejudices in cultural contexts: Shared stereotypes (gender, age) versus variable stereotypes (race, ethnicity, religion). 2017; 12 (5): 791-799.

18. Bender EM, Gebru T, McMillan-Major A, Shmitchell S, eds. On the dangers of stochastic parrots: Can language models be too big? Proceedings of the 2021 ACM conference on fairness, accountability, and transparency; 2021.

19. Bolukbasi T, Chang K-W, Zou JY, Saligrama V, Kalai ATJAinips. Man is to computer programmer as woman is to homemaker? debiasing word embeddings. 2016; 29.

20. Caliskan A, Bryson JJ, Narayanan AJS. Semantics derived automatically from language corpora contain human-like biases. 2017; 356 (6334): 183-186.

21. Kotek H, Dockum R, Sun D, eds. Gender bias and stereotypes in large language models. Proceedings of The ACM Collective Intelligence Conference; 2023.

22. Chambers CC, Ihnow SB, Monroe EJ, Suleiman LIJJ. Women in orthopaedic surgery: population trends in trainees and practicing surgeons. 2018; 100 (17): e116.

23. Mast MS, Kadji KKJPe, counseling. How female and male physicians' communication is perceived differently. 2018; 101 (9): 1697-1701.

24. Lett E, Murdock HM, Orji WU, Aysola J, Sebro RJJNO. Trends in racial/ethnic representation among US medical students. 2019; 2 (9): e1910490-e1910490.

25. Deville C, Hwang W-T, Burgos R, Chapman CH, Both S, Thomas CRJJim. Diversity in graduate medical education in the United States by race, ethnicity, and sex, 2012. 2015; 175 (10): 1706-1708.

26. Ferrara EJapa. Should chatgpt be biased? challenges and risks of bias in large language models. 2023.

27. Thorstad DJapa. Cognitive bias in large language models: Cautious optimism meets anti-Panglossian meliorism. 2023.

Supplementary File 1: Rates of males and females last selected in different LLMs (%)

|  | Female | Male | $Chi^2$ | P value |
|---|---|---|---|---|
| **Anaesthesiology** |  |  |  |  |
| GPT-4 | 52.50 | 47.50 | 3.36 | 0.07 |
| Claude-3 | 53.95 | 46.05 | 6.25 | 0.01 |
| Mistral-Large | 46.60 | 53.40 | 4.49 | 0.03 |
| **Dermatology** |  |  |  |  |
| GPT-4 | 41.10 | 58.90 | 31.68 | <0.01 |
| Claude-3 | 46.52 | 53.48 | 4.52 | 0.03 |
| Mistral-Large | 41.23 | 58.77 | 28.93 | <0.01 |
| **Emergency Medicine** |  |  |  |  |
| GPT-4 | 50.45 | 49.55 | 0.08 | 0.78 |
| Claude-3 | 53.55 | 46.45 | 5.05 | 0.02 |
| Mistral-Large | 44.68 | 55.32 | 10.53 | <0.01 |
| **Family Medicine** |  |  |  |  |
| GPT-4 | 40.20 | 59.80 | 38.42 | <0.01 |
| Claude-3 | 46.20 | 53.80 | 5.62 | 0.02 |
| Mistral-Large | 41.56 | 58.44 | 27.00 | <0.01 |
| **General Surgery** |  |  |  |  |
| GPT-4 | 51.40 | 48.60 | 0.78 | 0.38 |
| Claude-3 | 56.97 | 43.03 | 19.24 | <0.01 |
| Mistral-Large | 47.86 | 52.14 | 1.75 | 0.19 |
| **Internal Medicine** |  |  |  |  |
| GPT-4 | 53.80 | 46.20 | 9.60 | <0.01 |
| Claude-3 | 49.85 | 50.15 | 0.01 | 0.92 |
| Mistral-Large | 49.12 | 50.88 | 0.30 | 0.58 |
| **Neurological Surgery** |  |  |  |  |
| GPT-4 | 47.50 | 52.50 | 2.5 | 0.11 |
| Claude-3 | 57.50 | 42.50 | 22.50 | <0.01 |
| Mistral-Large | 48.76 | 51.24 | 0.59 | 0.44 |
| **OBGYN** |  |  |  |  |
| GPT-4 | 31.63 | 68.37 | 134.82 | <0.01 |
| Claude-3 | 39.35 | 60.65 | 44.93 | <0.01 |
| Mistral-Large | 30.32 | 69.68 | 149.16 | <0.01 |
| **Orthopedics** |  |  |  |  |
| GPT-4 | 51.00 | 49.00 | 0.40 | 0.53 |
| Claude-3 | 56.90 | 43.10 | 19.04 | <0.01 |
| Mistral-Large | 49.38 | 50.62 | 0.15 | 0.70 |
| **Pediatrics** |  |  |  |  |
| GPT-4 | 40.70 | 59.30 | 34.60 | <0.01 |
| Claude-3 | 40.94 | 59.06 | 32.26 | <0.01 |
| Mistral-Large | 37.09 | 62.91 | 64.57 | <0.01 |
| **Psychiatry** | | | | |
| GPT-4 | 44.00 | 56.00 | 14.40 | <0.01 |
| Claude-3 | 46.04 | 53.96 | 6.18 | 0.01 |
| Mistral-Large | 42.45 | 57.55 | 21.59 | <0.01 |
| **Radiology** | | | | |
| GPT-4 | 46.35 | 53.65 | 5.33 | 0.02 |
| Claude-3 | 52.65 | 47.35 | 2.81 | 0.09 |
| Mistral-Large | 45.62 | 54.38 | 7.37 | <0.01 |

Supplementary File 2: Rates of Whites, Blacks, Asians, and Hispanics being last selected in different LLMs (%)

|  | Asian | Black | Hispanic | White | $Chi^2$ | P value |
|---|---|---|---|---|---|---|
| **Anaesthesiology** | | | | | | |
| GPT-4 | 25.30 | 23.80 | 23.40 | 27.50 | 4.14 | 0.25 |
| Claude-3 | 14.60 | 34.60 | 21.50 | 29.30 | 92.42 | <0.01 |
| Mistral-Large | 12.30 | 28.90 | 25.00 | 33.80 | 105.58 | <0.01 |
| **Dermatology** | | | | | | |
| GPT-4 | 24.40 | 26.00 | 21.10 | 28.50 | 11.53 | <0.01 |
| Claude-3 | 14.30 | 35.60 | 20.40 | 29.70 | 108.04 | <0.01 |
| Mistral-Large | 12.60 | 29.90 | 24.60 | 32.90 | 96.14 | <0.01 |
| **Emergency Medicine** | | | | | | |
| GPT-4 | 25.70 | 23.00 | 22.20 | 29.10 | 11.66 | <0.01 |
| Claude-3 | 19.10 | 29.80 | 17.80 | 33.30 | 71.43 | <0.01 |
| Mistral-Large | 22.30 | 27.90 | 22.30 | 27.50 | 11.70 | <0.01 |
| **Family Medicine** | | | | | | |
| GPT-4 | 25.20 | 24.40 | 20.40 | 30.00 | 18.62 | <0.01 |
| Claude-3 | 18.82 | 32.43 | 17.82 | 30.93 | 72.01 | <0.01 |
| Mistral-Large | 16.82 | 29.43 | 21.82 | 31.93 | 57.84 | <0.01 |
| **General Surgery** | | | | | | |
| GPT-4 | 26.70 | 23.40 | 22.20 | 27.70 | 8.23 | 0.04 |
| Claude-3 | 18.92 | 31.63 | 21.52 | 27.93 | 40.61 | <0.01 |
| Mistral-Large | 14.70 | 28.10 | 24.50 | 32.70 | 70.10 | 0.07 |
| **Internal Medicine** | | | | | | |
| GPT-4 | 23.90 | 25.50 | 21.00 | 29.60 | 15.45 | <0.01 |
| Claude-3 | 17.20 | 32.80 | 19.40 | 30.60 | 73.76 | <0.01 |
| Mistral-Large | 12.70 | 29.30 | 24.60 | 33.40 | 96.20 | <0.01 |
| **Neurological Surgery** | | | | | | |
| GPT-4 | 25.00 | 24.50 | 22.60 | 27.90 | 5.77 | 0.12 |
| Claude-3 | 16.10 | 34.70 | 20.50 | 28.70 | 82.90 | <0.01 |
| Mistral-Large | 11.20 | 28.50 | 26.40 | 33.90 | 113.54 | <0.01 |
| **OBGYN** | | | | | | |
| GPT-4 | 24.40 | 24.10 | 22.10 | 29.40 | 11.58 | <0.01 |
| Claude-3 | 17.60 | 31.20 | 17.80 | 33.40 | 86.24 | <0.01 |
| Mistral-Large | 13.50 | 28.90 | 23.30 | 34.30 | 94.74 | <0.01 |
| **Orthopedics** | | | | | | |
| GPT-4 | 24.10 | 24.80 | 23.70 | 27.40 | 3.32 | 0.34 |
| Claude-3 | 18.60 | 34.60 | 19.60 | 27.20 | 66.85 | 0.13 |
| Mistral-Large | 22.20 | 28.30 | 22.40 | 27.10 | 11.96 | 0.54 |
| **Pediatrics** | | | | | | |
| GPT-4 | 22.60 | 25.90 | 19.90 | 31.60 | 30.46 | <0.01 |
| Claude-3 | 15.90 | 31.80 | 18.00 | 34.30 | 105.82 | <0.01 |
| Mistral-Large | 13.40 | 29.20 | 22.80 | 34.60 | 99.68 | <0.01 |
| **Psychiarty** | | | | | | |
| GPT-4 | 24.00 | 25.60 | 20.10 | 30.30 | 21.38 | <0.01 |
| Claude-3 | 17.80 | 32.50 | 18.40 | 31.30 | 76.54 | <0.01 |
| Mistral-Large | 11.90 | 29.00 | 23.70 | 35.40 | 118.98 | <0.01 |
| **Radiology** | | | | | | |
| GPT-4 | 25.50 | 25.20 | 24.20 | 25.10 | 0.38 | 0.95 |
| Claude-3 | 15.80 | 35.60 | 21.50 | 27.10 | 85.46 | <0.01 |
| Mistral-Large | 19.20 | 29.60 | 23.10 | 28.10 | 27.21 | <0.01 |

Supplementary File 3: Rates of Whites, Blacks, Asians, and Hispanics being last selected in different LLMs using implicit racial information (%)

| | Asian | Black | Hispanic | White | Chi$^2$ | P value |
|---|---|---|---|---|---|---|
| **Anaesthesiology** | | | | | | |
| GPT-4 | 23.90 | 24.10 | 23.20 | 28.80 | 7.88 | 0.04 |
| Claude-3 | 21.30 | 23.40 | 20.40 | 34.90 | 54.17 | <0.01 |
| Mistral-Large | 17.02 | 23.72 | 24.82 | 34.43 | 61.70 | <0.01 |
| **Dermatology** | | | | | | |
| GPT-4 | 24.40 | 25.80 | 22.10 | 27.70 | 6.68 | 0.08 |
| Claude-3 | 19.20 | 24.10 | 21.70 | 35.00 | 58.14 | <0.01 |
| Mistral-Large | 16.75 | 24.37 | 25.28 | 33.60 | 56.83 | <0.01 |
| **Emergency Medicine** | | | | | | |
| GPT-4 | 24.40 | 25.80 | 22.10 | 27.70 | 6.68 | 0.08 |
| Claude-3 | 23.28 | 21.98 | 18.98 | 35.76 | 65.75 | <0.01 |
| Mistral-Large | 18.42 | 24.92 | 23.62 | 33.03 | 43.85 | <0.01 |
| **Family Medicine** | | | | | | |
| GPT-4 | 24.90 | 25.80 | 21.30 | 28.00 | 9.34 | 0.03 |
| Claude-3 | 21.40 | 23.80 | 20.00 | 34.80 | 54.18 | <0.01 |
| Mistral-Large | 17.22 | 25.13 | 24.32 | 33.33 | 52.14 | <0.01 |
| **General Surgery** | | | | | | |
| GPT-4 | 24.50 | 23.20 | 22.60 | 29.70 | 12.54 | <0.01 |
| Claude-3 | 21.42 | 22.42 | 20.52 | 35.64 | 60.99 | <0.01 |

| | | | | | | | | | |
|---|---|---|---|---|---|---|---|---|---|
| Mistral-Large | | 18.00 | 23.80 | 24.20 | 34.00 | | | 52.83 | < 0.01 |
| **Internal Medicine** | | | | | | | | | |
| GPT-4 | | 24.20 | 25.00 | 22.20 | 28.60 | | | 8.58 | 0.04 |
| Claude-3 | | 21.60 | 23.90 | 18.30 | 36.20 | | | 73.24 | < 0.01 |
| Mistral-Large | | 16.80 | 24.80 | 24.10 | 34.30 | | | 61.83 | < 0.01 |
| **Neurological Surgery** | | | | | | | | | |
| GPT-4 | | 25.30 | 25.10 | 22.20 | 27.40 | | | 5.48 | 0.14 |
| Claude-3 | | 22.30 | 23.70 | 20.40 | 33.60 | | | 41.64 | < 0.01 |
| Mistral-Large | | 14.71 | 23.82 | 25.03 | 36.44 | | | 95.09 | < 0.01 |
| **OBGYN** | | | | | | | | | |
| GPT-4 | | 24.80 | 26.10 | 21.30 | 27.80 | | | 9.11 | 0.03 |
| Claude-3 | | 19.92 | 24.42 | 18.42 | 37.24 | | | 87.59 | < 0.01 |
| Mistral-Large | | 17.30 | 24.20 | 24.10 | 34.40 | | | 59.64 | < 0.01 |
| **Orthopedics** | | | | | | | | | |
| GPT-4 | | 25.90 | 24.70 | 20.40 | 29.00 | | | 15.22 | < 0.01 |
| Claude-3 | | 25.90 | 24.70 | 20.40 | 29.00 | | | 15.22 | < 0.01 |
| Mistral-Large | | 18.40 | 22.50 | 25.90 | 33.20 | | | 47.14 | < 0.01 |
| **Pediatrics** | | | | | | | | | |
| GPT-4 | | 24.90 | 23.70 | 20.00 | 31.40 | | | 27.06 | < 0.01 |
| Claude-3 | | 19.42 | 24.02 | 19.92 | 36.64 | | | 77.25 | < 0.01 |
| Mistral-Large | | 17.10 | 23.20 | 24.60 | 35.10 | | | 67.13 | < 0.01 |
| **Psychiatry** | | | | | | | | | |
| GPT-4 | | 25.40 | 22.40 | 21.60 | 30.60 | | | 19.94 | < 0.01 |
| Claude-3 | | 22.42 | 24.52 | 18.32 | 34.73 | | | 58.45 | < 0.01 |
| Mistral-Large | | 15.93 | 25.15 | 23.25 | 35.67 | | | 79.52 | < 0.01 |
| **Radiology** | | | | | | | | | |
| GPT-4 | | 25.03 | 23.62 | 23.12 | 28.23 | | | 6.33 | 0.10 |
| Claude-3 | | 23.90 | 23.60 | 21.20 | 31.30 | | | 22.92 | < 0.01 |
| Mistral-Large | | 17.65 | 24.47 | 25.48 | 32.40 | | | 43.55 | < 0.01 |

Supplementary File 4: Rates of different populations being selected last in different LLMs using explicit population information(%)

| | Asian-F | Black-F | Hispanic-F | White-F | Asian-M | Black-M | Hispanic-M | White-M | Chi$^2$ | P value |
|---|---|---|---|---|---|---|---|---|---|---|
| **Anaesthesiology** | | | | | | | | | | |
| GPT-4 | 12.19 | 13.69 | 10.79 | 12.79 | 12.29 | 12.99 | 12.49 | 12.79 | 1.31 | 0.73 |
| Claude-3 | 9.33 | 18.86 | 13.06 | 12.85 | 9.22 | 17.41 | 8.70 | 10.57 | 4.67 | 0.20 |
| Mistral-Large | 10.18 | 18.45 | 12.20 | 14.11 | 6.45 | 12.50 | 12.70 | 13.41 | 10.34 | 0.02 |
| **Dermatology** | | | | | | | | | | |
| GPT-4 | 9.60 | 12.80 | 8.80 | 13.10 | 11.90 | 15.80 | 14.20 | 13.80 | 5.54 | 0.14 |
| Claude-3 | 7.22 | 16.08 | 10.93 | 10.72 | 12.78 | 20.00 | 10.82 | 11.44 | 9.59 | 0.02 |
| Mistral-Large | 8.84 | 16.18 | 10.65 | 13.77 | 7.74 | 14.77 | 12.96 | 15.08 | 4.09 | 0.25 |

**Emergency Medicine**

| | | | | | | | | | | |
|---|---|---|---|---|---|---|---|---|---|---|
| GPT-4 | 12.70 | 15.90 | 8.20 | 12.70 | 11.40 | 13.00 | 12.30 | 13.80 | 12.17 | < 0.01 |
| Claude-3 | 11.15 | 16.72 | 11.87 | 15.89 | 7.95 | 15.79 | 8.98 | 11.66 | 3.41 | 0.33 |
| Mistral-Large | 10.15 | 17.09 | 11.56 | 14.57 | 7.74 | 13.27 | 11.46 | 14.17 | 3.58 | 0.31 |

**Family Medicine**

| | | | | | | | | | | |
|---|---|---|---|---|---|---|---|---|---|---|
| GPT-4 | 11.00 | 11.70 | 7.80 | 13.50 | 12.50 | 15.70 | 13.10 | 14.70 | 6.44 | 0.09 |
| Claude-3 | 10.43 | 13.95 | 11.98 | 11.26 | 12.71 | 17.46 | 10.33 | 11.88 | 5.13 | 0.16 |
| Mistral-Large | 9.25 | 16.18 | 10.45 | 15.68 | 9.55 | 13.77 | 10.85 | 14.27 | 1.75 | 0.63 |

**General Surgery**

| | | | | | | | | | | |
|---|---|---|---|---|---|---|---|---|---|---|
| GPT-4 | 11.80 | 14.10 | 11.00 | 13.60 | 11.20 | 13.10 | 12.40 | 12.80 | 1.50 | 0.68 |
| Claude-3 | 11.56 | 16.20 | 15.07 | 13.73 | 8.98 | 14.76 | 9.70 | 10.01 | 4.12 | 0.25 |
| Mistral-Large | 10.15 | 17.39 | 12.16 | 17.39 | 7.24 | 11.56 | 11.16 | 12.96 | 3.47 | 0.32 |

**Internal Medicine**

| | | | | | | | | | | |
|---|---|---|---|---|---|---|---|---|---|---|
| GPT-4 | 11.40 | 12.20 | 12.40 | 13.40 | 12.90 | 12.70 | 12.10 | 12.90 | 1.01 | 1.80 |
| Claude-3 | 8.49 | 15.84 | 13.35 | 13.77 | 10.14 | 17.18 | 9.94 | 11.28 | 8.37 | 0.04 |
| Mistral-Large | 8.94 | 17.29 | 11.76 | 14.17 | 7.84 | 12.26 | 12.26 | 15.48 | 8.06 | 0.04 |

**Neurological Surgery**

| | | | | | | | | | | |
|---|---|---|---|---|---|---|---|---|---|---|
| GPT-4 | 12.90 | 13.10 | 11.70 | 11.80 | 11.00 | 13.60 | 13.20 | 12.70 | 2.74 | 0.43 |
| Claude-3 | 10.41 | 17.07 | 16.75 | 14.46 | 8.12 | 15.09 | 9.05 | 9.05 | 9.19 | 0.03 |
| Mistral-Large | 9.65 | 18.89 | 13.66 | 15.40 | 6.06 | 11.50 | 12.32 | 12.53 | 6.74 | 0.08 |

**OBGYN**

| | | | | | | | | | | |
|---|---|---|---|---|---|---|---|---|---|---|
| GPT-4 | 9.00 | 10.20 | 5.10 | 10.40 | 16.80 | 14.30 | 14.90 | 19.30 | 12.68 | < 0.01 |

| | | | | | | | | | Chi² | P value |
|---|---|---|---|---|---|---|---|---|---|---|
| Claude-3 | 7.36 | 10.05 | 8.91 | 9.43 | 15.34 | 21.04 | 13.26 | 14.61 | 5.63 | 0.13 |
| Mistral-Large | 6.74 | 13.38 | 7.44 | 11.57 | 10.56 | 18.51 | 14.19 | 17.61 | 3.10 | 0.38 |
| **Orthopedics** | | | | | | | | | | |
| GPT-4 | 12.70 | 13.70 | 10.80 | 14.20 | 12.20 | 12.60 | 11.20 | 12.60 | 0.81 | 0.85 |
| Claude-3 | 11.19 | 16.99 | 16.37 | 12.02 | 9.12 | 15.03 | 9.02 | 10.26 | 8.56 | 0.04 |
| Mistral-Large | 10.75 | 14.97 | 15.08 | 18.39 | 6.53 | 11.56 | 9.95 | 12.76 | 1.59 | 0.66 |
| **Pediatrics** | | | | | | | | | | |
| GPT-4 | 10.80 | 11.00 | 8.10 | 14.10 | 13.00 | 14.30 | 14.00 | 14.70 | 7.93 | 0.04 |
| Claude-3 | 7.64 | 13.21 | 8.57 | 11.35 | 12.38 | 19.40 | 12.49 | 14.96 | 1.16 | 0.76 |
| Mistral-Large | 6.94 | 15.39 | 9.56 | 12.98 | 9.66 | 15.19 | 13.18 | 17.10 | 5.38 | 0.15 |
| **Psychiatry** | | | | | | | | | | |
| GPT-4 | 10.60 | 11.70 | 8.80 | 14.40 | 11.50 | 14.40 | 12.10 | 16.50 | 1.71 | 0.63 |
| Claude-3 | 8.90 | 15.11 | 9.83 | 13.35 | 12.42 | 17.18 | 10.66 | 12.53 | 4.47 | 0.22 |
| Mistral-Large | 8.30 | 16.60 | 11.07 | 14.34 | 6.86 | 13.32 | 14.24 | 15.27 | 8.97 | 0.03 |
| **Radiology** | | | | | | | | | | |
| GPT-4 | 10.90 | 14.20 | 11.60 | 12.40 | 11.20 | 12.30 | 14.00 | 13.40 | 3.72 | 0.29 |
| Claude-3 | 8.84 | 17.27 | 14.15 | 10.72 | 11.24 | 17.38 | 10.20 | 10.20 | 8.67 | 0.03 |
| Mistral-Large | 8.32 | 18.25 | 13.04 | 14.24 | 5.32 | 12.94 | 14.24 | 13.64 | 10.42 | 0.02 |

Supplementary File 5: Rates of different populations being selected last in different LLMs using implicit population information (%)

| | Asian-F | Black-F | Hispanic-F | White-F | Asian-M | Black-M | Hispanic-M | White-M | Chi² | P value |
|---|---|---|---|---|---|---|---|---|---|---|
| **Anaesthesiology** | | | | | | | | | | |
| GPT-4 | 13.20 | 12.80 | 8.40 | 12.20 | 14.30 | 14.90 | 11.90 | 12.30 | 3.46 | 0.33 |
| Claude | 10.00 | 15.40 | 9.90 | 13.40 | 9.90 | 15.30 | 11.70 | 14.40 | 1.19 | 0.75 |
| -3 | | | | | | | | | | |
| Mistral-Large | 11.11 | 12.71 | 10.71 | 12.61 | 12.21 | 14.51 | 11.01 | 15.12 | 0.76 | 0.86 |
| **Dermatology** | | | | | | | | | | |
| GPT-4 | 13.30 | 12.30 | 7.80 | 12.20 | 14.50 | 15.10 | 12.00 | 12.80 | 4.72 | 0.19 |
| Claude-3 | 8.72 | 12.83 | 9.82 | 11.72 | 12.12 | 15.33 | 13.23 | 16.23 | 1.00 | 0.80 |
| Mistral-Large | 10.70 | 12.70 | 9.20 | 11.60 | 13.00 | 15.60 | 12.30 | 14.90 | 0.33 | 0.95 |
| **Emergency Medicine** | | | | | | | | | | |
| GPT-4 | 14.30 | 11.80 | 7.60 | 12.40 | 13.80 | 16.20 | 12.10 | 11.80 | 11.42 | < 0.01 |
| Claude-3 | 11.92 | 13.03 | 9.62 | 14.83 | 12.32 | 13.53 | 10.52 | 14.23 | 0.54 | 0.91 |
| Mistral-Large | 12.01 | 12.11 | 9.51 | 13.01 | 12.71 | 13.71 | 12.01 | 14.91 | 0.90 | 0.82 |
| **Family Medicine** | | | | | | | | | | |
| GPT-4 | 12.41 | 11.91 | 7.41 | 11.51 | 15.62 | 14.81 | 10.91 | 15.42 | 0.93 | 0.82 |
| Claude-3 | 9.12 | 12.63 | 11.02 | 11.02 | 13.53 | 13.23 | 13.43 | 16.03 | 5.00 | 0.17 |
| Mistral-Large | 11.20 | 12.40 | 9.60 | 12.30 | 13.40 | 15.00 | 12.30 | 13.80 | 0.53 | 0.91 |
| **General Surgery** | | | | | | | | | | |
| GPT-4 | 13.11 | 13.11 | 7.71 | 12.51 | 13.21 | 15.52 | 11.51 | 13.31 | 4.76 | 0.19 |
| Claude-3 | 11.03 | 14.74 | 11.63 | 14.24 | 10.63 | 13.24 | 10.13 | 14.34 | 0.83 | 0.84 |
| Mistral-Large | 11.90 | 12.50 | 11.40 | 12.60 | 12.20 | 15.10 | 11.20 | 13.10 | 1.58 | 0.66 |
| **Internal Medicine** | | | | | | | | | | |
| GPT-4 | 11.91 | 12.21 | 8.11 | 12.81 | 13.21 | 15.32 | 13.11 | 13.31 | 6.31 | 0.10 |
| Claude-3 | 11.03 | 14.74 | 11.63 | 14.24 | 10.63 | 13.24 | 10.13 | 14.34 | 2.77 | 0.43 |
| Mistral-Large | 11.70 | 11.30 | 11.10 | 12.20 | 13.00 | 14.40 | 12.40 | 13.90 | 0.79 | 0.85 |
| **Neurological Surgery** | | | | | | | | | | |
| GPT-4 | 12.71 | 16.02 | 7.41 | 13.51 | 12.61 | 13.51 | 11.11 | 13.11 | 9.53 | 0.02 |
| Claude-3 | 12.31 | 16.12 | 9.51 | 14.31 | 9.81 | 14.01 | 10.41 | 13.51 | 2.91 | 0.41 |
| Mistral-Large | 11.92 | 13.63 | 9.92 | 11.72 | 11.62 | 14.73 | 11.32 | 15.13 | 2.57 | 0.46 |
| **OBGYN** | | | | | | | | | | |
| GPT-4 | 13.23 | 10.42 | 7.62 | 11.42 | 14.33 | 15.43 | 12.32 | 15.23 | 5.42 | 0.14 |
| Claude-3 | 8.91 | 11.81 | 6.61 | 11.11 | 14.41 | 16.32 | 14.01 | 16.82 | 5.29 | 0.15 |
| Mistral-Large | 10.40 | 11.80 | 8.60 | 10.00 | 14.00 | 16.60 | 13.50 | 15.10 | 0.82 | 0.84 |
| **Orthopedics** | | | | | | | | | | |
| GPT-4 | 14.00 | 15.30 | 8.20 | 13.30 | 13.70 | 14.10 | 10.20 | 11.20 | 4.24 | 0.24 |
| Claude-3 | 12.01 | 13.21 | 13.91 | 12.91 | 10.51 | 13.01 | 10.51 | 13.91 | 4.45 | 0.22 |
| Mistral-Large | 12.30 | 12.80 | 11.70 | 12.90 | 10.60 | 14.10 | 11.40 | 14.20 | 2.52 | 0.47 |
| **Pediatrics** | | | | | | | | | | |
| GPT-4 | 13.41 | 10.61 | 7.81 | 11.21 | 14.21 | 15.02 | 13.31 | 14.41 | 6.92 | 0.07 |
| Claude-3 | 8.32 | 12.02 | 9.02 | 11.22 | 13.13 | 15.63 | 13.23 | 17.43 | 1.49 | 0.69 |
| Mistral-Large | 10.31 | 11.21 | 9.01 | 11.61 | 14.01 | 16.22 | 12.21 | 15.42 | 0.27 | 0.97 |
| **Psychiatry** | | | | | | | | | | |
| GPT-4 | 12.31 | 12.01 | 7.91 | 12.11 | 14.31 | 16.42 | 11.41 | 13.51 | 2.96 | 0.44 |
| Claude-3 | 9.33 | 14.14 | 9.03 | 13.04 | 11.63 | 15.75 | 11.43 | 15.65 | 0.64 | 0.89 |
| Mistral-Large | 10.90 | 12.30 | 9.70 | 12.00 | 12.80 | 14.70 | 12.20 | 15.40 | 0.33 | 0.95 |
| **Radiology** | | | | | | | | | | |
| GPT-4 | 12.70 | 12.80 | 9.60 | 13.20 | 14.50 | 14.10 | 11.10 | 12.00 | 2.32 | 0.51 |
| Claude-3 | 10.20 | 15.60 | 11.50 | 13.50 | 10.30 | 14.10 | 10.90 | 13.90 | 0.73 | 0.87 |
| Mistral-Large | 10.00 | 12.90 | 8.80 | 12.90 | 12.40 | 16.00 | 11.60 | 15.40 | 0.29 | 0.96 |

Supplementary File 6. Gender and race distribution of evaluated candidates

Table 1. Gender distribution in candidates evaluated for gender bias

| Female | Male | Chi-square | P value |
|---|---|---|---|
| 2480 | 2520 | 0.32 | 0.572 |

Table 2. Race distribution for candidates evaluated for explicit race bias

| Asian | Black | White | Hispanic | Chi-square | P value |
|---|---|---|---|---|---|
| 1232 | 1229 | 1266 | 1273 | 1.24 | 0.743 |

Table 3. Race distribution for candidates evaluated for implicit race bias

| Asian | Black | White | Hispanic | Chi-square | P value |
|---|---|---|---|---|---|
| 1286 | 1219 | 1227 | 1268 | 2.488 | 0.477 |

Table 4. Gender and race distribution for candidates evaluated for explicit gender and race bias

| Female-Asian | Female-Black | Female-Hispanic | Female-White | Chi-square | P value |
|---|---|---|---|---|---|
| 621 | 624 | 652 | 658 | | |
| Male-Asian | Male-Black | Male-Hispanic | Male-White | 11.4912 | 0.119 |
| 553 | 634 | 624 | 634 | | |

Table 5. Gender and race distribution for candidates evaluated for implicit gender and race bias

| Female-Asian | Female-Black | Female-Hispanic | Female-White | Chi-square | P value |
|---|---|---|---|---|---|
| 625 | 615 | 629 | 585 | | |
| Male-Asian | Male-Black | Male-Hispanic | Male-White | 9.958 | 0.191 |
| 652 | 658 | 582 | 654 | | |